\documentclass[pra,amsmath,amssymb, aps, reprint, longbibliography,nofootinbib,nobibnotes]{revtex4-2}
\usepackage[utf8]{inputenc}
\usepackage[T1]{fontenc}

\usepackage{amsthm}
\usepackage{amsmath,bm,bbm}
\usepackage{amssymb}
\usepackage{amsfonts}
\usepackage{braket}
\usepackage{graphicx}
\usepackage{mathtools}
\usepackage[unicode=true, breaklinks=false, pdfborder={0 0 1}, backref=false, colorlinks=true, linkcolor=blue, citecolor=blue, urlcolor=blue]{hyperref}

\usepackage{dsfont}

\usepackage{graphicx} 

\hypersetup{citecolor=magenta}
\hypersetup{colorlinks=true}
\hypersetup{linkcolor=blue}
\hypersetup{urlcolor=blue}

\newcommand{\mat}[1]{\begin{pmatrix} #1 \end{pmatrix}}

\DeclareMathOperator\tr{tr}

\newcommand{\da}{\dagger}
\newcommand{\la}{\langle}
\newcommand{\ra}{\rangle}
\newcommand{\id}{\ensuremath{\mathbbm 1}}

\newcommand{\reel}{\mathbb{R}}
\newcommand{\erw}{\mathbb{E}}
\newcommand{\cD}{\mathcal{D}}
\newcommand{\diff}{\mathrm{d}}
\newcommand{\e}{\mathrm{e}}
\renewcommand{\vec}[1]{\boldsymbol{#1}}
\newcommand{\Op}[1]{\hat{#1}}
\newcommand{\vx}{\vec{x}}
\newcommand{\vy}{\vec{y}}
\newcommand{\vz}{\vec{z}}
\newcommand{\vX}{\vec{X}}

\newcommand{\vp}{\vec{p}}
\newcommand{\vq}{\vec{q}}
\newcommand{\ve}{\vec{e}}
\newcommand{\vP}{\vec{P}}
\newcommand{\vQ}{\vec{Q}}
\newcommand{\vR}{\vec{R}}

\newcommand{\vv}{\vec{v}}
\newcommand{\vw}{\vec{w}}

\newcommand{\ox}{\Op{x}}
\newcommand{\op}{\Op{p}}
\newcommand{\oa}{\Op{a}}
\newcommand{\oH}{\Op{H}}
\newcommand{\oM}{\Op{M}}
\newcommand{\oL}{\Op{L}}
\newcommand{\oell}{\Op{\ell}}

\newcommand{\oK}{\Op{K}}
\newcommand{\oA}{\Op{A}}

\newcommand{\oS}{\Op{S}}
\newcommand{\oB}{\Op{B}}
\newcommand{\oU}{\Op{U}}
\newcommand{\oW}{\Op{W}}
\newcommand{\ovx}{\Op{\vec{x}}}
\newcommand{\ovp}{\Op{\vec{p}}}

\newcommand{\opi}{\Op{\pi}}
\newcommand{\oxi}{\Op{\xi}}
\newcommand{\vf}{\vec{f}}
\newcommand{\vF}{\vec{F}}
\newcommand{\Hess}{\mathbb{H}}
\newcommand{\trans}{\mathrm{T}}

\begin{document}
\title{Quantum measurement feedback models of friction beyond the diffusive limit and their connection to collapse models}

\author{Michael Gaida}
\email{michael.gaida@student.uni-siegen.de}
\affiliation{Naturwissenschaftlich--Technische Fakult\"{a}t, Universit\"{a}t Siegen, 
Walter-Flex-Stra{\ss}e 3, 57068 Siegen, Germany}
\author{Stefan Nimmrichter}
\email{stefan.nimmrichter@uni-siegen.de}
\affiliation{Naturwissenschaftlich--Technische Fakult\"{a}t, Universit\"{a}t Siegen, 
Walter-Flex-Stra{\ss}e 3, 57068 Siegen, Germany}

\begin{abstract}
 We present and discuss a master equation blueprint for a generic class of quantum measurement-feedback based models of friction. A desired velocity-dependent friction force is realized on average by random repeated applications of unsharp momentum measurements followed by immediate outcome-dependent momentum displacements. The master equations can describe arbitrarily strong measurement-feedback processes as well as the weak continuous limit resembling diffusion master equations of Caldeira-Leggett type. 
 We show that the special case of linear friction can be equivalently represented by an average over random position measurements with squeezing and position displacements as feedback. In fact, the dissipative continuous spontaneous localization model of objective wavefunction collapse realizes this representation for a single quantum particle. We reformulate a consistent many-particle generalization of this model and highlight the possibility of feedback-induced correlations between otherwise non-interacting particles.
\end{abstract}

\maketitle

\section{Introduction} 

Friction is a ubiquitous phenomenon that affects most of the dynamics we observe and use in everyday life. It is often described phenomenologically as an emergent force between macroscopic objects and media, conveniently kept under wraps in undergraduate lectures on analytical mechanics. Microscopic quantum models of friction are typically based on two paradigms: Mediated interactions between mechanical systems that come with retardation or delay, which lead to velocity dependent forces; and the coupling of mechanical systems to a large thermal reservoir, which converts useful motion into uncontrolled excitations of many environmental degrees of freedom.

An example of the first paradigm is the so-called quantum friction that naturally arises as a velocity dependent force between polarizable objects in relative motion \cite{JBPendry_1997,buhmann-2012}. An example of the second paradigm is the Caldeira-Leggett master equation describing the quantum Brownian motion of a system, arising either from continuous weak coupling to a bath of harmonic oscillators \cite{CALDEIRA1983587}, or from repeated collisions with a background gas \cite{ QLB}.

In contrast, many friction forces in the lab are engineered by means of external control in order to cool down motional degrees of freedom close to their ground state---a crucial prerequisite for many quantum experiments with cold atoms \cite{Metcalf2012} or in  optomechanics \cite{doi:10.1126/science.aba3993,Millen_2020}. For example, if the internal structure of the system is accessible, Doppler cooling can be used to dissipate kinetic energy \cite{HANSCH197568} via spontaneous emission. Alternatively, a loss of kinetic energy can also be achieved by coherent light scattering into a damped cavity mode \cite{horak1997cavity}. In feedback cooling, on the other hand, one monitors the particle position and applies an appropriate feedback force \cite{rossi2018measurement, tebbenjohanns2020motional}. 

The latter feedback approaches to cooling typically operate in the linearized, diffusive regime of continuous weak measurements and weak Markovian feedback, for which the closed-loop system dynamics are well described using Langevin equations or Caldeira-Leggett-like master equations \cite{wiseman_milburn_2009}. Here, we take one step back and consider a generic blueprint of Markovian measurement-feedback friction models based on the random repeated application of finite-resolution momentum measurements followed by finite-strength unitary feedback operations. While a similar idea was already formulated for position measurements and worked out in the diffusive regime in \cite{PhysRevA.36.5543}, we explore explicitly non-diffusive models for friction forces based on momentum measurements---with generalized diffusion master equations appearing as a limiting case.

After introducing our model and its main predicitons for a single-particle state in Section \ref{sec_measurement_based_models}, we focus on the most prominent and convenient special case of linear friction in Section \ref{sec:linearFriction}.
We prove that its outcome-averaged closed-loop dynamics can be equivalently generated by position measurements followed by feedback operations comprised of squeezing and position-dependent translations. This will generally lead to a non-thermal equilibrium state, which we will investigate in an exemplary harmonic oscillator case study.

The last part of our paper, Section \ref{sec:dCSL}, explores the relation between our framework and spontaneous collapse models, which are objective modifications to the Schrödinger equation that reinstate macroscopic realism by gradually localizing the wave function of massive systems \cite{bassi2013models,GRW,Diosi,Hilbert_SDE}. These models violate energy conservation, and recent experiments have placed stringent bounds on the model parameters \cite{carlesso_present_2022,sym15020480}, putting forth variations of these models that could include dissipation. 
We focus on the dissipative extension of the most-studied Continuous Spontaneous Localization model (dCSL) \cite{DCSL}, which restricts the unbounded heating predicted by the original CSL model \cite{CSL,Hilbert_SDE} to a finite energy scale. We find that the ensemble-averaged dynamics of a single particle under dCSL can be understood as a measurement-feedback friction process, i.e., falls under our generic master equation model for linear friction. 
Finally, we show that, unline CSL, the many-particle generalization of the dCSL model cannot be based on an arbitrary reference mass scale and may induce correlations between non-interacting particles in close vicinity.

\section{Measurement-feedback friction models}
\label{sec_measurement_based_models}

We formulate a blueprint model for measurement-feedback friction of a single particle based on general (possibly non-Gaussian) momentum measurements followed immediately by feedback momentum kicks, which occur randomly in time at a chosen rate. 

Consider a probability density $\mu(\vp) \geq 0$ over three-dimensional momenta $\vp$ and let $ \oM (\vq) = \sqrt{\mu (\ovp - \vq)} $ be the associated POVM operators, so that
\begin{equation}
    \int \diff^3 \vq \, \oM^\da (\vq) \oM(\vq) = \int \diff^3 \vq \, \mu (\vq) \id = \id .
\end{equation}
Formally, such a POVM can be thought of as arising from the idealized pointer interaction with an appropriately initialized ancilla state \cite{PhysRevA.36.5543,barchielli_model_1982}, where $\mu$ is usually taken to be Gaussian. 
Also, the measurement operators could in principle be based on any complex square root of $\mu$, i.e., in all the following, $\sqrt{\mu(\vq)} = e^{i\varphi(\vq)} |\sqrt{\mu(\vq)}|$ may be understood more generally as a complex-valued function whose absolute square is $\mu(\vq)$. 

Notice that the unconditional outcome-averaged measurement has the same net effect on the particle state as a random unitary process of position displacements \cite{vacchini2007precise},
\begin{equation}
    \int \diff^3 \vq \ \oM(\vq) \rho \oM^\dagger (\vq) = \int \diff^3 \vy \ \nu (\vy) \ \e^{i \vy \cdot \ovp/\hbar} \rho \ \e^{-i \vy \cdot \ovp / \hbar} \ . \label{eq_random_trans}
\end{equation}
The displacements are distributed according to the even probability density
\begin{equation}
    \nu (\vy) = \nu (-\vy) = \frac{1}{(2\pi\hbar)^3} \left| \int \diff^3 \vq\,  \e^{ i \vq \cdot \vy/\hbar} \sqrt{\mu(\vq)} \right|^2, \label{eq:nu}
\end{equation}
normalized by virtue of Plancherel's theorem. The identity \eqref{eq_random_trans}  proves helpful in later calculations.

Coherent feedback without delay can be implemented by applying a conditional unitary $\oU(\vq)$ to the post-measurement state, as described by the modified POVM operators $\oL(\vq) = \oU(\vq) \oM(\vq)$ and the conditional nonlinear state transformations
\begin{equation}
    \rho \mapsto \Lambda_{\vq} [\rho] = \frac{\oL(\vq)\rho\oL^\da(\vq)}{\tr \{ \oL^\da \oL(\vq) \rho \} } = \frac{\oU(\vq) \oM(\vq) \rho \oM(\vq) \oU^\da (\vq)}{\tr \{ \oM^2 (\vq) \rho \}} .
\end{equation}
We restrict our view to momentum kicks as feedback, 
\begin{equation}
    \oU (\vq) = \e^{-i \vf(\vq) \cdot \ovx / \hbar},
\end{equation}
with a chosen momentum kick by $-\vf(\vq)$, conditioned on the measurement outcome $\vq$. This kick could be realized by rapid controlled collisions or force impulses and shall implement a desired mean friction force. 
The unconditional state transformation, averaged over all outcomes, is then given by the linear CPTP map
\begin{equation}
    \Lambda \left[ \rho \right] = \int \diff^3 \vq \ \oU (\vq) \oM(\vq) \ \rho \ \oM(\vq) \oU ^\dagger (\vq) \ . \label{eq:Lambda_channel}
\end{equation}
It is easy to see that this channel is translation-covariant,
\begin{equation}
    \Lambda \left[ \e^{i \vz \cdot \ovp / \hbar} \rho \ \e^{-i \vz \cdot \ovp / \hbar} \right] = 
     \e^{i \vz \cdot \ovp / \hbar}  \Lambda \left[  \rho \right] \e^{-i \vz \cdot \ovp/\hbar}
\end{equation}
and thus does not depend on the chosen origin of the particle coordinate system. However, the channel does depend on the chosen inertial reference frame as the feedback unitary will single out the origin of momentum as the rest velocity.

For a dynamical friction model, we assume that the prescribed measurement-feedback process continuously affects the particle state at an average rate $\Gamma$, either in the form of a Poisson process of repeated jump events or gradually building up continuously in time in the form of a  diffusion process. Ensemble-averaged over all possible random trajectories, we can describe the evolution of the particle state $\rho_t$ over a small time step from $t$ to $t+\diff t$ as follows. With probability $1-\Gamma \diff t$, the particle state evolves undisturbed and unitarily according to its Hamiltonian $\oH$, whereas with probability $\Gamma \diff t \ll 1$, it is subjected to the channel \eqref{eq:Lambda_channel}, so that
\begin{align}
    \rho_{t + \diff t} =& \left( 1 - \Gamma \diff t \right) \e^{-i \oH  \diff t/\hbar} \rho_t \   \e^{i \oH  \diff t/\hbar} \nonumber \\
    &+ \Gamma \diff t \  \Lambda \left[ \rho_t \right] + \mathcal{O}\left( \diff t^2 \right) 
\end{align}
In the limit $\diff t \to 0$, we can expand the coherent part and arrive at the Markovian master equation
\begin{equation}
    \frac{\diff}{\diff t} \rho_t = - \frac{i}{\hbar} \left[ \oH , \rho_t \right] + \Gamma \left( \Lambda \left[ \rho_t \right] - \rho_t \right) \ , \label{eq_master_general}
\end{equation}
with its Lindblad operators given by $\oL(\vq)$. This master equation is a special case of the continuous monitoring ansatz \cite{Hornberger_2007}, which can also be understood as the closed-loop master equation for a coherent feedback control model based on efficient measurements \cite{wiseman_milburn_2009}.

\subsection{Friction and diffusion effect}

The time evolution of the mean and the variance of the particle's position and momentum provides a clear picture of the impact of the measurement-feedback master equation \eqref{eq_master_general}. For any observable $\oA$, the expectation value $\la \oA \ra_t = \tr \{ \oA \rho_t \}$ evolves according to
\begin{equation}\label{eq:Ehrenfest_eq_A}
    \frac{\diff}{\diff t} \left\langle \oA  \right\rangle_t = \frac{i}{\hbar} \left\langle \left[ \oH , \oA  \right] \right\rangle_t + \Gamma  \left\langle \Lambda^\dagger [ \oA  ] \right\rangle_t - \Gamma  \left\langle \oA  \right\rangle_t \ .
\end{equation}
Here, $\Lambda^\da$ denotes the adjoint channel defined through $\tr\{ \Lambda^\da [\oA] \rho\} \equiv \tr\{ \oA \Lambda[\rho]\}$, which has the form 
\begin{equation}
    \Lambda^\dagger \left[ \oA \right] = \int \diff \vq \ \oM(\vq) \oU^\dagger(\vq) \oA \oU(\vq) \oM(\vq). \label{eq:Lambda_adj}
\end{equation}

Denoting classical expectation values by $\erw_{\mu(\vq)} \left[ f(\vq) \right] = \int \diff^3 \vq \ \mu(\vq) f(\vq) $, we can express the transformation of purely momentum- or position-dependent functions under the adjoint channel by
\begin{align}
     \Lambda^\dagger \left[F \left( \ovp \right) \right]   &= \erw_{\mu(\vq)} \left[ F \left( \ovp - \vf \left( \ovp - \vq \right) \right) \right] \  , \label{eq:Lambda_F} \\
     \Lambda^\dagger \left[ G \left( \ovx \right) \right]  &=  \erw_{\nu(\vy)} \left[  G \left( \ovx - \vy \right) \right] \  . \label{eq:Lambda_G}
\end{align}
See Appendix \ref{App_adj_channel} for a step-by-step derivation. Note that, in case of the position operator, we have made use of \eqref{eq_random_trans}, and the result does not depend on the choice of the feedback function $\vf$. 

From the above expressions and the fact that $\nu$ is an even, zero-mean distribution, we can deduce the time evolution of the first and second moments of position and momentum. The mean values evolve according to
\begin{align}
    \frac{\diff }{\diff t} \left\langle \ovx \right\rangle_t &= \frac{i}{\hbar} \left\la [\oH,\ovx] \right\ra_t \, , \label{eq:dxdt_general} \\
    \frac{\diff }{\diff t} \left\langle \ovp \right\rangle_t &= \frac{i}{\hbar} \left\la [\oH,\ovp] \right\ra_t - \Gamma \, \erw_{\mu(\vq)} \left[ \left\langle \vf \left( \ovp - \vq \right)  \right\rangle_t \right] \, . \label{eq:dpdt_general}
\end{align}
Leaving the Hamiltonian part aside, we observe that the measurement-feedback channel leaves the mean particle position unaffected, while inducing a momentum-dependent mean force, as determined by the chosen feedback function $\vf$ smeared out over the distribution of uncertain measurement outcomes. 

For the pure second moments, we have 
\begin{align}
    \frac{\diff }{\diff t} \left\langle \ovx^2 \right\rangle_t &= \frac{i}{\hbar} \left\la [\oH,\ovx^2] \right\ra_t + \Gamma \erw_{\nu(\vy)} \left[ \vy^2 \right] \, , \label{eq:dx2dt_general} \\ \nonumber
    \frac{\diff }{\diff t} \left\langle \ovp^2 \right\rangle_t &= \frac{i}{\hbar} \left\la [\oH,\ovp^2] \right\ra_t 
    - 2\Gamma \left\langle \ovp \cdot \erw_{\mu(\vq)} \left[ \vf(\ovp-\vq) \right] \right\rangle \\ 
    &+ \Gamma \left\langle \erw_{\mu(\vq)} \left[ \vf^2(\ovp - \vq) \right] \right\rangle . \label{eq:dp2dt_general}
\end{align}
Here, we observe that the momentum measurement causes position diffusion, i.e., a constant growth of the position uncertainty over time. The momentum uncertainty, on the other hand, may be lowered by the effective friction force, but this is compromised by the fact that the feedback is in itself noisy as it is based on uncertain measurement outcomes.

We remark that $\sqrt{\mu}$ and $\sqrt{\nu}$ are related to each other by Fourier transform, in the same fashion as position and momentum wave functions. This implies an uncertainty relation between the second moments,
\begin{equation}
    \left( \erw_{\mu(\vq)} \left[ \vq^2 \right] -  \erw_{\mu(\vq)} \left[ \vq \right]^2 \right) \erw_{\nu(\vy)} \left[ \vy^2 \right]
    \geq \frac{3}{4} \hbar^2,
\end{equation}
which expresses a trade-off between the feedback noise and the measurement noise that the process imparts on the particle in \eqref{eq:dx2dt_general} and \eqref{eq:dp2dt_general}. 

In order to realize a net friction force that slows down the particle, the feedback kick $-\vf$ should point opposite to the measured momentum, i.e., $\vf(\vq) = \bar{f}(q) \vq/q $ with some $f(q) \geq 0$. Together with the additional requirement of an isotropic measurement distribution $\mu (\vq) = \bar{\mu} (q)$, this ensures that the measurement-feedback channel $\Lambda$ is rotationally covariant; see Appendix \ref{App_Rot_cov} for a proof.
The simplest and most natural choices for the feedback include: (i) $\vf(\vq) = \alpha \vq$ mimicking linear friction with a dimensionless strength parameter $\alpha>0$, (ii) $\vf(\vq) = \alpha_C \vq/q$ mimicking a constant friction, and (iii) $\vf (\vq) = \alpha_S q \vq$ mimicking a quadratic drag. We will restrict our view on the linear option (i), as it keeps calculations comparably simple and admits a straightforward many-particle generalization. The other two options could be considered as viable alternatives that result in a relative enhancement of the friction rate at (ii) smaller or (iii) greater velocities.

\subsection{Diffusion limit and the Caldeira-Leggett master equation}\label{sec:diffusionLimit}

The majority of case studies in quantum feedback control and its quantum optical implementations are concerned with the limit of weak measurement-feedback processes, mostly in the form of a continuous diffusion process. In our model \eqref{eq_master_general}, this amounts to a broad measurement distribution $\mu$ that barely resolves the particle's momentum state and to small momentum kicks $\vf$ that barely displace the particle. Expanding the measurement-feedback channel up to second order in the position and momentum quadratures leaves us with a diffusion master equation,
\begin{align}
\label{eq_pre_CD}
    \frac{\diff}{\diff t} \rho  &\approx  - \frac{i}{\hbar} \left[ \oH - \vF \cdot \ovx, \rho \right] - \Gamma \sum_{jk} \Big( 
    A_{jk} \left[ \op_j, \left[ \op_k, \rho \right] \right] \\  
    &+ iB_{jk} \left[ \ox_j, \left\{ \op_k, \rho \right\} \right] + C_{jk} \left[ \ox_j, \left[ \ox_k, \rho \right] \right] \Big); \nonumber
\end{align}
see Appendix \ref{App_diff_limit} for a detailed derivation. 
Here $\vF = \Gamma \erw_\mu [\vf]$ denotes a constant mean drift force and $A,B,C$ are real-valued diffusion matrices with elements
\begin{align}\label{eq:diffABC}
    A_{jk} &= -\frac{1}{8} \erw_{\mu(\vq)} \left[ \frac{\partial^2 \ln \mu(\vq)}{\partial q_j \partial q_k} \right] = \frac{\erw_{\nu(\vy)} [y_j y_k ]}{2 \hbar^2}, \\ 
    B_{jk} &= \frac{1}{2 \hbar} \erw_{\mu(\vq)} \left[ \frac{\partial f_j(\vq)}{\partial q_k} \right], \quad 
    C_{jk} = \frac{\erw_{\mu(\vq)} [f_j(\vq) f_k(\vq) ]}{2 \hbar^2} . \nonumber 
  \end{align}
  Leaving the Hamiltonian $\oH$ aside, these terms can be identified as the drift and diffusion coefficients of a six-dimensional Fokker-Planck equation that evolves the Wigner function representing the particle state in phase space \cite{barnett_methods_2002}. The matrices $A$ and $C$ are responsible for position and momentum diffusion, respectively, while $B$ incorporates an effective friction force.

Conversely, if one is given a friction-diffusion master equation of the above form \eqref{eq_pre_CD}, the identities \eqref{eq:diffABC} can be viewed as prescriptions for constructing a measurement distribution $\mu$ and feedback function $\vf$ such that \eqref{eq_pre_CD} approximates the master equation \eqref{eq_master_general}. Contrary to the former, the latter is comprised of mathematically benign bounded measurement and unitary feedback operators, which can improve numerical stability in simulations.

As an instructive example that we will expand upon in the following sections, consider the case of linear friction, $\vf(\vq) = \alpha\vq$, and a Gaussian measurement distribution with zero mean and  covariance matrix $\Sigma \geq 0$, 
\begin{equation}
    \mu (\vq) = \frac{1}{(2\pi)^{-3/2} \sqrt{\det(\Sigma)}} \e^{- \frac{1}{2} \vq^\trans \Sigma^{-1} \vq}.
\end{equation}
This choice results in a vanishing drift, $\vF = 0$, while the diffusion matrices simplify to
\begin{equation}
    A = \frac{1}{8}\Sigma^{-1}, \quad B = \frac{\alpha}{2\hbar} \id, \quad C =  \frac{\alpha^2}{2\hbar^2} \Sigma .
\end{equation}
They are all diagonal in the orthonormal eigenbasis of $\Sigma$. Without loss of generality, we can thus take these eigenvectors as the Cartesian axes of our coordinate system, which allows us to express the diffusion master equation \eqref{eq_pre_CD} in diagonal form,
\begin{align} \label{eq:ME_diffusionLimit}
    \frac{\diff}{\diff t} \rho = &- \frac{i}{\hbar} \left[ \oH, \rho \right] - \frac{\Gamma}{2} \sum_{j=1}^3 \Big(  \frac{1}{4 \sigma_j^2} \left[ \op_j, \left[ \op_j, \rho \right] \right]  \\ \nonumber
    &+ i \frac{\alpha}{\hbar} \left[ \ox_j, \left\{ \op_j, \rho \right\} \right] + \frac{\alpha^2 \sigma_j^2}{2 \hbar^2}  \left[ \ox_j \left[ \ox_j, \rho \right] \right]    \Big), 
\end{align}

where the $\sigma_j^2$ denote the eigenvalues of $\Sigma$. This form highlights the tradeoff between measurement and feedback noise in the process. Outcome-averaged momentum measurements lead to position diffusion in proportion to $1/\sigma_j^2$, which is less invasive the poorer the measurement resolution. Conversely, the feedback is based on the measurement outcomes and therefore inflicts the more momentum diffusion the greater the measurement uncertainties $\sigma_j^2$. 

The master equation \eqref{eq:ME_diffusionLimit} can be rewritten in the familiar Lindblad form \cite{breuer_petruccione},
\begin{align}\label{eq:CL_ME}
    \frac{\diff}{\diff t} \rho = &- \frac{i}{\hbar} \left[ \oH + \frac{\Gamma \alpha}{2} \sum_{j=1}^3 \left\{ \ox_j, \op_j \right\}, \rho \right] + \frac{\Gamma}{2} \sum_{j=1}^3 \cD [\oell_j]\rho,
\end{align}
with the usual abbreviation $\cD[\oL]\rho = \oL\rho\oL^\da - \{\oL^\da\oL,\rho\}/2$ and the Lindblad operators 
\begin{equation}
    \oell_j = \alpha \frac{\sqrt{2} \sigma_j}{\hbar} \ox_j + i \frac{1}{ \sqrt{2} \sigma_j} \op_j.
\end{equation}
We recognize \eqref{eq:CL_ME} as a three dimensional and completely positive version of the Caldeira-Legett master equation \cite{CALDEIRA1983587,breuer_petruccione}.

\section{Single-particle models with linear friction force}\label{sec:linearFriction}

We now put our focus on measurement-feedback models for linear friction on a single particle, viz.~the channel \eqref{eq:Lambda_channel} and the master equation \eqref{eq_master_general} with $\vf(\vq) = \alpha \vq$. 
We show that the same channel can be realized in a different manner, as a position measurement-feedback protocol. Moreover, we discuss the dissipation dynamics under the master equation, highlighting that the steady state is in general not a Gibbs state. As an exemplary case study, we consider one-dimensional motion in a harmonic trap. Finally, we comment on the advantages of restricting to the often used Gaussian measurements.

\subsection{Equivalence to a position measurement-feedback channel}

\begin{figure*}
    \centering
 \includegraphics[width=\textwidth]{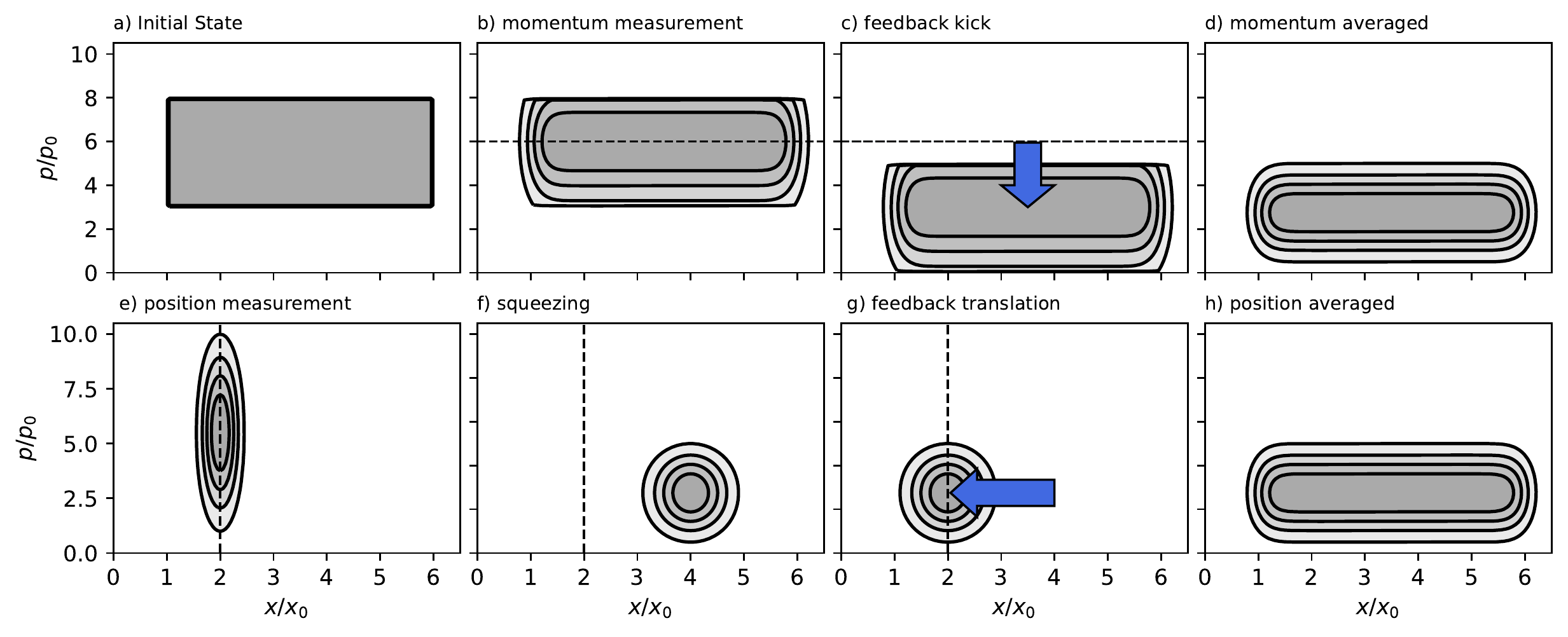}
    \caption{Illustration of two alternative quantum measurement-feedback protocols representing the same linear friction channel $\Lambda$ of strength $\alpha=1/2$ in one-dimensional phase space. Positions and momenta are given in some natural units $x_0$ and $p_0$. Starting from the exemplary Wigner function in (a), the first protocol consists in (b) an unsharp momentum measurement, here with outcome $q=6p_0$, followed by (c) a momentum displacement by $-\alpha q = -3p_0$, which after averaging over all possible outcomes $q$ yields the final state (d). Alternatively, one performs (e) a conjugate unsharp position measurement with outcome $y=2x_0$, followed by (f) a momentum squeezing operation by $\xi=-\ln 2$ and (g) a translation by $\alpha y/(\alpha-1) = -2x_0$, which results in the same outcome-averaged state (h) as before.}
    \label{fig_Wigner}
\end{figure*}

The Kraus representation of a quantum channel is not unique. In the present case of a continuous set of Kraus operators $\oL(\vq)$, consider an invertible transformation to the new set of operators,
\begin{equation}\label{eq:KfromL}
    \oK (\vy) = \int \diff^3 \vq \, u(\vy,\vq) \oL(\vq),
\end{equation}
where the kernel obeys $\int \diff^3 \vy \, u(\vy,\vq) u^{*} (\vy,\vq') = \delta (\vq-\vq')$. It is then clear that
\begin{equation}\label{eq:Lambda_L_K}
    \Lambda [\rho] = \int\diff^3\vq\, \oL(\vq)\rho\oL^\da (\vq) = \int\diff^3\vy\, \oK(\vy) \rho \oK^\da (\vy).
\end{equation}
We obtain a surprisingly simple and intuitive alternative representation of the measurement-feedback channel \eqref{eq:Lambda_channel} for linear friction $\vf(\vq) = \alpha \vq$ if we choose the rescaled Fourier kernel $u(\vy,\vq) = (\alpha/2\pi\hbar)^{3/2} \exp (i\alpha \vy\cdot\vq/\hbar)$ and restrict to $0 < \alpha < 1$: 
\begin{align}
    \oK(\vy) =& \left(\frac{\alpha}{2\pi\hbar} \right)^{3/2} \int \diff^3 \vq \, e^{i[\alpha \vy\cdot\vq-\vf(\vq)\cdot \ovx]/\hbar} \sqrt{\mu(\ovp-\vq)} \nonumber \\
    =& \exp \left( \frac{i\ovp}{\hbar} \cdot \frac{\alpha \vy}{1-\alpha}\right) \oS(\alpha) \nonumber \\
    & \times \sqrt{\left(\frac{\alpha}{1-\alpha}\right)^3 \nu \left[ \frac{\alpha}{1-\alpha} (\ovx-\vy)  \right]} \label{eq:Ky}
\end{align}
In other words, the channel $\Lambda$ comprised of unconditional momentum measurements according to the distribution $\mu$ with outcomes $\vq$ and subsequent linear momentum translations by $-\alpha \vq$ can be equivalently represented by position measurements according to the conjugate distribution \eqref{eq:nu} over positions rescaled by $\alpha/(1-\alpha)$, followed by unitary feedback. The latter consists in the (outcome-independent) squeezing operation 
\begin{equation}
    \oS (\alpha) = 
    \exp \left[ i \ln \left(1-\alpha\right)\frac{ \ovx\cdot\ovp + \ovp\cdot\ovx}{2 \hbar} \right] , \label{eq:Salpha}
\end{equation}
followed by a linear position translation by $-\alpha \vy/(1-\alpha)$. The proof of this equivalence is detailed (and extended to $\alpha >1$) in Appendix~\ref{App_mom_to_pos}. A pictorial representation for the one-dimensional case is sketched in Fig.~\ref{fig_Wigner}. 

Once again, we take $\sqrt{\nu}$ as a generalized (possibly complex-valued) square-root of $\nu$ defined via
\begin{equation}\label{eq:sqrt_nu}
    \sqrt{\nu(\vy)} = \int \frac{\diff^3 \vq}{ (2 \pi \hbar)^{3/2}} \e^{i \vq \cdot \vy / \hbar} \sqrt{\mu(\vq)}.
\end{equation}
The squeezing caused by \eqref{eq:Salpha} can be made explicit by introducing auxiliary harmonic frequencies $\omega_j$ and bosonic operators $\oa_j$ for the Cartesian components $j=1,2,3$, such that $\oa_j = \sqrt{m \omega_j/2 \hbar} ( \ox_j + i \op_j/m\omega  )$. We can then express the unitary \eqref{eq:Salpha} as a product of three single-mode squeezing operators defined in the usual manner \cite{gerry_knight},
\begin{equation}\label{eq:S_squeezingForm}
    \oS(\alpha) = \bigotimes_{j=1}^3 \exp \left[ \frac{\xi_j^* \oa_j^2 - \xi_j \oa_j^{\da 2}}{2} \right],
\end{equation}
with squeezing parameters $\xi_j = \ln(1-\alpha)$.

We notice that the Kraus operators \eqref{eq:Ky} are ill-defined for $\alpha=0$, which shows that the equivalence between the two channels only holds in the presence of a feedback force. For $\alpha=1$ and $\alpha>1$, we have similar but rather technical results that are explained in Appendix \ref{App_mom_to_pos}. 

The representation of the measurement-feedback channel in terms of the $\oK(\vy)$ will be useful in Sec.~\ref{sec_singleDCSL} where we show that the dCSL model for a single particle is equivalent to the linear measurement-feedback friction model with a Gaussian distribution $\mu$.

\subsection{Dissipation effect and steady state}

Let the particle be subject to random applications of the measurement-feedback channel with rate $\Gamma$ in an arbitrary potential, $\oH = \ovp^2/2m + V(\ovx)$.
At first glance, the linear feedback induces a net slowing of the particle for any $\alpha >0$, as apparent from the resulting damping term in the equation of motion for the average momentum \eqref{eq:dpdt_general},
\begin{equation}
    \frac{\diff}{\diff t} \left\la  \ovp \right\ra_t = - \left\la \nabla V(\ovx) \right\ra_t - \Gamma \alpha \left\la \ovp \right\ra_t + \Gamma \alpha \erw_{\mu (\vq)} [\vq]. \label{eq:dpdt_linear}
\end{equation}
Presuming that the momentum measurement is unbiased, $\erw_{\mu (\vq)} [\vq] = 0$, a free particle would eventually come to a halt on average, $\la \ovp\ra_{t\to\infty} \to 0$.
However, damping of average momentum does not always imply dissipation of kinetic energy. Its time evolution is given by 
\begin{align}
   \frac{\diff}{\diff t} \left\la  \ovp^2 \right\ra_t =& - \left\la \ovp\cdot \nabla V(\ovx) + h.c. \right\ra_t -\Gamma \alpha (2-\alpha) \la \ovp^2 \ra_t \nonumber \\
   &- 2\Gamma \alpha (1-\alpha) \la \ovp \ra_t \cdot \erw_{\mu (\vq)} [\vq] + \Gamma \alpha^2 \erw_{\mu(\vq)} [\vq^2],   \label{eq:dp2dt_linear}
\end{align}
which results in a damping term only for $0<\alpha<2$. Indeed, the linear feedback function $\vf$ describes a translation by $\alpha$ times the outcome $\vq$ of an imprecise measurement of the particle momentum. In the absence of bias ($\erw_{\mu (\vq)} [\vq] = 0$), the possible outcomes would be centered around $\la \ovp\ra_t$, and kicking the particle by more than twice the magnitude of that outcome would thus result in average heating rather than dissipation. 

The last term in \eqref{eq:dp2dt_linear} contributes momentum diffusion at a constant rate, which ultimately limits how much kinetic energy can be dissipated. This feedback noise is due to the imprecise momentum measurement; it would vanish in the limit of a perfect measurement, $\mu(\vq) = \delta (\vq)$, which however implies diverging measurement noise, i.e., position diffusion in \eqref{eq:dx2dt_general}.

In the friction regime $\alpha \in (0,2)$, we expect the particle to reach a finite-energy steady state. In general, this state will not be a Gibbs state of thermal equilibrium, $\rho_{\rm th} = e^{-\oH/k_B T}/Z$, with one notable exception that we will highlight shortly. The reason is intuitively clear: For the Gibbs (or any other $\oH$-diagonal) state to be a steady state, it must also be a fix point of the measurement feedback channel $\Lambda$. Since $\Lambda$ is translationally covariant and causes unbounded position diffusion, we conclude that a unique steady state would have to be translationally invariant, which cannot be $\oH$-diagonal unless $V(\ovx)=\text{const}\equiv 0$.

More formally, let 
\begin{equation}
    \chi_\rho (\vP,\vX) = \tr \left\{ e^{i(\vP\cdot\ovx + \ovp\cdot\vX)/\hbar} \rho \right\}
\end{equation}
be the Wigner-Weyl characteristic function representation of the single-particle state $\rho$ \cite{breuer_petruccione}, i.e., the Fourier transform of its Wigner function. A straightforward calculation reveals that the linear friction channel $\Lambda$ transforms it like 
\begin{align}
    \chi_{\Lambda[\rho]} (\vP,\vX) &= A(\vP,\vX) e^{- i \alpha \vP\cdot\vX/2\hbar} \chi_\rho (\vP,(1-\alpha)\vX), \nonumber \\
    A(\vP,\vX) &= \int\diff^3\vq \, e^{i\alpha \vq\cdot\vX/\hbar} \sqrt{\mu(\vq-\vP)\mu(\vq)}. \label{eq:charFunc_trafo}
\end{align}
Hence, a fix point of $\Lambda$ must obey
\begin{equation}
    \chi_\rho (\vP,(1-\alpha)\vX) \stackrel{!}{=} \frac{e^{-i \alpha \vP\cdot\vX/2\hbar}}{A(\vP,\vX)} \chi_\rho (\vP,\vX). \label{eq:fixPointCond}
\end{equation}
We prove in Appendix \ref{App_trafo_char} that this can only hold for singular functions, such as functions of the form $\chi_\rho (\vP,\vX) = \delta(\vP) \xi (\vX)$ that correspond to translation-invariant improper states.

For the special case of a free particle, Gibbs states correspond to singular characteristic functions of the above form, with $\xi (\vX) = e^{-mk_B T \vX^2/2\hbar^2}$. Setting $\vP=0$ in \eqref{eq:fixPointCond} then leads to 
\begin{equation}
    A(0,\vX) \stackrel{!}{=} e^{-mk_B T \vX^2 \alpha (2-\alpha)/2\hbar^2},
\end{equation}
which by virtue of \eqref{eq:charFunc_trafo} requires that the measurement distribution be an isotropic Gaussian, $\mu (\vq) = e^{-\vq^2/2\sigma^2}/(2\pi\sigma^2)^{3/2}$. Otherwise, the fix point will be stationary, but not a Gibbs state. 
We conclude that an isotropic Gaussian momentum measurement of resolution $\sigma$ with linear feedback of strength $\alpha$ thermalizes a single free particle to a temperature $k_B T = \alpha\sigma^2/(2-\alpha)m $.

\subsection{The case for an isotropic Gaussian measurement distribution}

While the friction feedback always distinguishes an inertial reference frame with zero velocity, we have seen that the measurement-feedback channel is still translation-covariant. For linear friction, or more generally for any central friction feedback $\vf(\vq) = \bar{f}(q) \vq/q$, one additionally achieves covariance under fixed rotations of the reference frame if and only if the measurement distribution is isotropic, $\mu(\vq) = \bar{\mu}(q)$. The proof is given in Appendix \ref{App_Rot_cov}.

Most of the literature considers POVM measurements of position or momentum that are exclusively Gaussian, typically following an indirect pointer measurement model \cite{PhysRevA.36.5543,barchielli_model_1982}. From our perspective, a crucial property of a Gaussian measurement distribution is that it factorizes, $\mu (\vq) = \mu_1(q_1) \mu_2(q_2) \mu_3(q_3)$. The same then holds for the distribution $\nu(\vy)$ in \eqref{eq:nu}. This ensures that the measurement channel with linear feedback does not build up directional correlations in the form of covariances between different momentum or position components. Indeed, we find that the adjoint channel $\Lambda^\da$ acts upon the mixed second moments as
\begin{align}
    \Lambda^\da [\op_j \op_k] =& (1-\alpha)^2\op_j \op_k + \alpha (1-\alpha) \Big( \erw_{\mu(\vq)} [q_k] \op_j \nonumber \\  
    &+ \erw_{\mu(\vq)} [q_j] \op_k \Big) + \alpha^2 \erw_{\mu(\vq)} [q_j q_k]  , \\
    \Lambda^\da [\ox_j \ox_k] =& \ox_j \ox_k + \erw_{\nu(\vy)} \left[y_j y_k \right] .
\end{align}
Accordingly, the Ehrenfest equations of motion under a dynamical application of the channel at rate $\Gamma$ read as
\begin{align}
    \frac{\diff}{\diff t} \la \op_j \op_k \ra_t &= - \left\la \op_j \partial_k V(\ovx) + \partial_j V(\ovx) \op_k \right\ra_t \nonumber \\ & \ \ \ \ + \alpha (1-\alpha) \Big( \erw_{\mu(\vq)} [q_k] \op_j  
    + \erw_{\mu(\vq)} [q_j] \op_k \Big)  \nonumber \\  
    & \ \ \ \ - \alpha(2-\alpha) \Gamma \left\la \op_j \op_k \right\ra_t + \alpha^2 \erw_{\mu(\vq)} [q_j q_k]  , \\
    \frac{\diff}{\diff t} \la \ox_j \ox_k \ra_t &= \frac{\la \ox_j \op_k + \op_j \ox_k \ra_t}{m} + \Gamma \erw_{\nu(\vy)} \left[ y_j y_k \right].
\end{align}
The constant input provided by the last terms in both equations would affect both the variances ($j=k$) and the covariances, but a factorizing, unbiased distribution $\mu$ only increases the variances. In fact, the Kraus operators themselves factorize into $\oL(\vq) = \bigotimes_j \oL_j (q_j)$ with $\oL_j (q_j) = e^{-i\alpha q_j \ox_j} \sqrt{\mu_j(\op_j-q_j)}$. Hence one can conveniently trace over the motion along any one or two of the spatial directions and obtain reduced channels of the same form acting individually on the reduced states, 
\begin{align}
    \Lambda_{1,2} [\rho_{1,2}] &= \tr_3 \{ \Lambda [\rho] \} \label{eq:Lambda12} \\
    &= \int \diff q_1 \diff q_2 \, \oL_1(q_1) \oL_2(q_2) \rho_{1,2}  \oL^\da_1(q_1) \oL^\da_2(q_2), \nonumber \\
    \Lambda_1 [\rho_1] &= \tr_{2,3} \{ \Lambda [\rho] \} = \int \diff q_1\, \oL_1(q_1) \rho_1 \oL^\da_1(q_1). \label{eq:Lambda1}
\end{align}
Both features together, translation covariance and factorization of the Kraus operators, would be realized by an isotropic and factorizing distribution function, 
\begin{equation}
    \mu(\vq) = \bar{\mu}\left(\sqrt{q_1^2 + q_2^2 + q_3^2} \right) = \mu_1(q_1) \mu_2(q_2) \mu_3(q_3). \label{eq:mu_isotropic_factorizing}
\end{equation}
It is easy to show that such a distribution can only be Gaussian: Evaluating \eqref{eq:mu_isotropic_factorizing} at the origin and on each coordinate axis ($\vq=q \ve_j$) yields 
$\mu_j (q)/\mu_j(0) = \bar{\mu} (|q|)/\bar{\mu}(0) \equiv \epsilon (q^2)$ for every $j=1,2,3$ and $q \in\reel$. In terms of the auxiliary function $\epsilon$, the condition \eqref{eq:mu_isotropic_factorizing} reduces to
\begin{equation}
    \epsilon(q_1^2 + q_2^2 + q_3^2) = \epsilon (q_1^2) \epsilon (q_2^2) \epsilon (q_3^2) \quad \forall q_{1,2,3} \in \reel,
\end{equation}
which is a functional equation constituting the exponential function. Hence, if we also demand a finite norm, we get $\bar{\mu}(q) \propto \epsilon(q^2) = e^{-q^2/2\sigma^2}$ with some standard deviation parameter $\sigma$.

\subsection{Equilibration of a harmonic oscillator} \label{subsec_HO}

As a simple case study, we consider the one-dimensional motion of a harmonically trapped particle, $\oH = \op^2/2m + m \omega^2 \ox^2/2$. Assuming a factorizing, but otherwise generic, momentum measurement distribution, we can ignore the other spatial dimensions and describe the measurement-feedback channel by the reduced map \eqref{eq:Lambda1}, with Kraus operators $\oL_1(q) = e^{-i\alpha q\ox/\hbar} \sqrt{\mu_1 (\op-q)}$ implementing linear friction.

The time evolution \eqref{eq_master_general} under random repeated applications of the channel yields a closed set of linear inhomogeneous Ehrenfest equations of motion for the first and second moments of $\ox$ and $\op$. For simplicity, we express these equations in terms of the dimensionless quadratures $\oxi = \ox/x_0$ and $\opi = \op/p_0$, where $x_0 = \sqrt{\hbar/m\omega}$ is the ground-state amplitude and $p_0 = \hbar/x_0$.

The first moments evolve according to 
\begin{equation}
    \frac{\diff}{\diff t} \mat{\la \oxi\ra_t \\ \la \opi\ra_t} = \mat{0 & \omega \\ -\omega & -\Gamma \alpha} \mat{\la \oxi\ra_t \\ \la \opi\ra_t} + \frac{\Gamma \alpha}{p_0}\mat{0 \\  \erw_{\mu_1(q)} [q] },
\end{equation}
which for $\alpha>0$ resembles a damped harmonic oscillator driven by a constant force proportional to the bias $\erw_{\mu_1(q)} [q]$. The oscillator is driven towards a stationary mean displacement of $\la \oxi\ra_{\infty} = \Gamma \alpha \erw_{\mu_1(q)}[q]/ \omega p_0$ over the damping time scale $1/\Gamma \alpha$.

\begin{figure}
    \centering
 \includegraphics[width=\columnwidth]{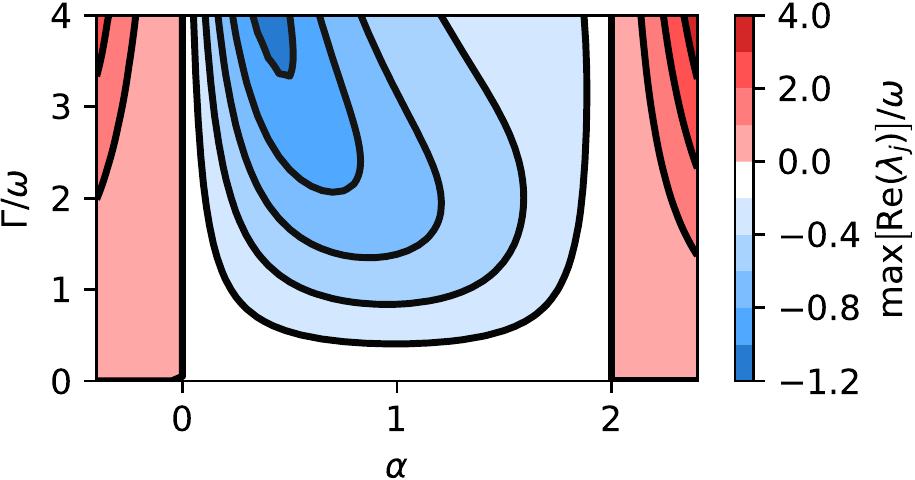}
    \caption{Greatest real part of the eigenvalues of the system matrix $S/\omega$ from \eqref{eq:S} for one-dimensional harmonic motion of frequency $\omega$ subject to repeated momentum measurements at the rate $\Gamma$ and linear feedback of strength $\alpha$. Negative values (blue) imply that all second moments of position and momentum converge to a stable equilibrium configuration of finite energy. This is the case for any $\Gamma>0$ and $\alpha \in (0,2)$.} 
    \label{fig_eigvals}
\end{figure}

For the second moments, we obtain the system 
\begin{align}
\label{eq_HO_second}
    &\frac{\diff}{\diff t} \mat{\la \oxi^2\ra_t \\ \la \{ \oxi,\opi \} \ra_t \\ \la \opi^2\ra_t} = S \mat{\la \oxi^2\ra_t \\ \la \{ \oxi,\opi \} \ra_t \\ \la \opi^2\ra_t} \\  &+ \Gamma \mat{ \erw_{\nu_1(y)} [y^2]/x_0^2 \\ 2  \alpha \la \oxi\ra_t \erw_{\mu_1(q)}[q]/ p_0 \\ \alpha^2 \erw_{\mu_1(q)} [q^2]/p_0^2 -2 \alpha(1-\alpha) \erw_{\mu_1(q)}[q] \la \opi \ra_t /p_0}, \nonumber
\end{align}
 or in short $\dot{\vv}_t = S \vv_t + \vw_t$, with the abbreviated column vectors $\vv_t,\vw_t$. Notice the coupling to the first moments via $\vw_t$ in case of a biased distribution. The system matrix reads as 
\begin{equation}\label{eq:S}
    S = \mat{  0 & \omega & 0 \\ -2\omega & -\Gamma \alpha & 2\omega \\ 0 & -\omega & -\Gamma \alpha (2-\alpha) }.
\end{equation}
It is invertible if $\det (S) = -2 \omega^2 \Gamma \alpha (2-\alpha) \neq 0$, in which case there is a unique equilibrium of second moments, $\vv_{\infty} = -S^{-1} \vw_{\infty}$. Whether this equilibrium is a stable attractor depends on the eigenvalues of $S$. We plot their greatest real part in Fig.~\ref{fig_eigvals} as a function of $\Gamma/\omega$ and $\alpha$. As expected, it remains negative for $0 < \alpha < 2$, resulting in a stable equilibrium at finite energy. Any other $\alpha$-value leads to an infinite heating of the motional state (provided, obviously, that $\Gamma >0$).

The equilibrium energy can be given explicitly as
\begin{align} \label{eq:HO_meanEnergyEq}
    \frac{ \la \oH \ra_\infty}{\hbar \omega} =& \frac{\alpha}{2 - \alpha} \frac{\erw_{\mu_1(q)} \left[ q^2 \right]}{p_0^2} + \frac{\alpha^2}{2} \left( \frac{\Gamma}{\omega} \right)^2 \left(\frac{\erw_{\mu_1(q)} \left[ q \right]}{p_0} \right)^2 \nonumber \\ 
    &+ \left[ \frac{1}{\alpha(2-\alpha)} + \frac{\alpha}{4} \left( \frac{\Gamma}{\omega} \right)^2 \right] \frac{\erw_{\nu_1(y)} \left[ y^2 \right]}{x_0^2}.
\end{align}
Although one might be tempted to associate an effective temperature $T$ to this equilibrium value, the equilibrium state is not a Gibbs state, $\rho_{\infty} \neq e^{-\oH/k_B T}/Z$. This is already evident from the equilibrium second moments,

\begin{align}
    \left\la \left\{ \oxi, \opi \right\} \right\ra_{\infty} &= - \left( \frac{\Gamma}{\omega} \right) \frac{\erw_{\nu_1(y)} \left[ y^2 \right]}{x_0^2}, \label{eq_Ho_corr} \\ \label{eq_Ho_equi} 
    \la \oxi^2  -  \opi^2 \ra_\infty &= \left( \frac{\Gamma}{\omega} \right)^2 \left[ \frac{\alpha \erw_{\nu_1(y)} \left[ y^2 \right]}{2x_0^2} +  \left(\frac{ \alpha \erw_{\mu_1(q)} \left[ q \right] }{p_0} \right)^2  \right]. 
\end{align}
The first term describes correlations between the position and the momentum quadrature, while the second term describes an imbalance between the position and the momentum spread. Both terms would exactly vanish for a Gibbs state, which in phase space corresponds to an isotropic Gaussian Wigner function. Decreasing the feedback strength $\alpha$ leads to a lower imbalance at a higher mean energy, but it leaves the correlation term unaffected. Only in the limit of slow measurements, $\Gamma \ll \omega$, does the equilibrium approximate a thermal state. To lowest order in $\Gamma/\omega$, the mean energy \eqref{eq:HO_meanEnergyEq} then becomes rate-independent,
\begin{equation}
    \frac{ \la \oH \ra_\infty}{\hbar \omega} \approx
    \frac{1}{2-\alpha} \left( \frac{\alpha\erw_{\mu_1(q)} \left[ q^2 \right]}{p_0^2} + \frac{\erw_{\nu_1(y)} \left[ y^2 \right]}{\alpha x_0^2} \right)
\end{equation}
as in the case of conventional weak-coupling equilibration with a thermal environment.

\section{Dissipative Continuous Spontaneous Localization} \label{sec:dCSL}

After laying out the blueprint for generic Markovian measurement-feedback based friction models, we now discuss its application in the context of macrorealistic collapse models. A prime example is the continuous spontaneous localization (CSL) model \cite{CSL}, which averts quantum superpositions and reinstates classical realism on the macroscale through a random process that gradually localizes the wave function of large masses by measuring the positions of its constituent particles. The unavoidable side effect is that an otherwise isolated system of particles would heat up indefinitely at a constant rate \cite{Hilbert_SDE}. To mitigate this effect, Ref.~\cite{DCSL} proposed a modified variant, the dissipative CSL (dCSL) model. The employed method to include dissipation draws from master equation models for collisional decoherence due to background gas scattering \cite{QLB}.

In the following, we show that the observable consequences of the single-particle dCSL model are equivalent to our measurement-feedback master equation for linear friction. We then construct a scale-invariant and exchange-symmetric many-particle generalization of the dCSL master equation based on the consistency principles stated in Ref.~\cite{Nimmrichter_2013}, which differs from the originally proposed many-particle formulation in \cite{DCSL} and reveals a crucial shortcoming of dCSL. 

\subsection{Single-particle description}
\label{sec_singleDCSL}

The dCSL model in \cite{DCSL} predicts that the ensemble-averaged state of a single particle of mass $m$ subject to the Hamiltonian $\oH$ evolves according to the master equation 
\begin{align}\label{eq:dCSL_ME_single}
    \frac{\diff}{\diff t} \rho_t =& -\frac{i}{\hbar} [\oH,\rho_t] + \frac{\gamma}{m_0^2} \int\diff^3 \vy \, \cD [\oB(\vy)] \rho, \\
    \oB(\vy) =& \frac{m}{(2 \pi \hbar)^3} \int \diff^3 \vQ \ \e^{i \vQ \cdot \left( \ovx - \vy \right)/\hbar} \e^{-\frac{r_{\mathrm{CSL}}^2}{2 \hbar^2} \vert (1+k)\vQ + 2k \ovp \vert^2} .
\end{align}
Here, $m_0$ is a reference mass unit (typically the proton mass), $r_{\rm CSL}$ a length scale parameter, $k$ a dimensionless dissipation strength parameter, and $\gamma$ a rate density parameter in units of volume per time. The unmodified CSL model corresponds to $k=0$, in which case one can rewrite $\gamma = (4 \pi r_{\mathrm{CSL}}^2)^{3/2} \lambda_{\rm CSL}$ in terms of the more common rate parameter $\lambda_{\rm CSL}$ representing the localization rate for a single proton. 

With help of several manipulations, one can show that the dCSL Lindblad operators $\oB (\vy)$ have the same form as the Kraus operators \eqref{eq:Ky} representing a measurement-feedback model with a Gaussian measurement distribution and linear friction,
\begin{equation}
    \mu(\vq) = \left( \frac{2kr_{\rm CSL}}{\sqrt{\pi}\hbar} \right)^3 e^{-(2 k r_{\rm CSL} q/\hbar)^2}, \quad \alpha = \frac{2k}{1+k}. \label{eq:dCSL_mu_alpha}
\end{equation}
Using the measurement rate
\begin{equation}
    \Gamma = \left( \frac{m}{m_0} \right)^2 \frac{\gamma}{\left[ 2 \sqrt{\pi} (1+k) r_{\rm CSL} \right]^3},
    \label{eq:dCSL_Gamma}
\end{equation}
we can then identify 
\begin{align} \label{eq:dCSL_K}
    \sqrt{\Gamma} \oK(\vy) =& \frac{\sqrt{\Gamma}}{\sqrt{2\pi \hbar}^3} \left( \frac{2k}{1+k} \right)^{3/2} \int \diff^3 \vq \, \e^{2ik (\vy - \ovx) \cdot \vq/\hbar(1+k)} \nonumber \\
    & \times \left( \frac{2  k r_{\rm CSL}}{\sqrt{ \pi} \hbar } \right)^{3/2} \e^{-\frac{r_{\rm CSL}^2}{2 \hbar^2} \left( 2k \ovp - 2k \vq \right)^2} \nonumber \\ 
    =& \sqrt{\Gamma} \left[ \frac{(1+k)r_{\rm CSL}}{2\pi \sqrt{\pi} \hbar^2} \right]^{3/2} \int \diff^3 \vQ \ 
    \e^{i \vQ \cdot (\ovx - \vy )/\hbar} \nonumber \\ 
    &\times \e^{- r_{\rm CSL}^2\left[ (1+k) \vQ + 2k \ovp \right]^2/2\hbar^2} = \frac{\sqrt{\gamma}}{m_0}  \oB(\vy), 
\end{align}
where we substituted $\vQ(\vq) = - 2k\vq/ (1+k)$.
Hence the single-particle dCSL master equation \eqref{eq:dCSL_ME_single} is a special case of our measurement-feedback model \eqref{eq_master_general}. 

The equivalence suggests alternative interpretations and unravellings of the dCSL master equation into stochastic Schrödinger equations \cite{breuer_petruccione}. The dCSL model itself unravels \eqref{eq:dCSL_ME_single} directly as a Gaussian diffusion process based on the Lindblad operators $\oB(\vy)$. Instead, we may also unravel the master equation as a piecewise deterministic or diffusive process based on our momentum measurement and feedback operators $\oL(\vq)$, which could prove useful and more robust in numerical simulations.

\subsection{Consistent many-particle generalization}\label{sec:many_part}

Macrorealistic collapse models are universal modifications to the Schrödinger equation that should apply consistently to many-body systems on arbitrary scales, in such a way that the collapse effect is negligible in the atomic mass and size regime, but quasi-instantaneous in the macro-world. This requires that the single-particle formulation be generalized to $N$-particle systems of arbitrary masses $m_n$ under the following symmetry and consistency constraints \cite{Nimmrichter_2013}: (i) the exchange symmetry between identical particles and most fundamental symmetries underlying nonrelativistic mechanics should be preserved, (ii) the reduced state of any $N'<N$ masses must be consistently described by the corresponding $N'$-version of the model, (iii) the center-of-mass motion of a rigid, point-like compound of $N$ masses must be described by the single-particle model at mass $M=\sum_{n=1}^N m_n$, and (iv) the effective localization rate should amplify with the total mass involved in a quantum superposition state.

Conditions (ii) and (iii), in particular, establish a local and scale-invariant theory: we can restrict our view to  the interrogated quantum particles only, and we are allowed to speak of point-like test masses, coarse-graining over rigidly bound constituents whenever their internal motion cannot be resolved. Violating the conditions implies that additional masses in close vicinity to the interrogated system can no longer be omitted as 'innocent bystanders' and that the motional state of a compound must be expanded in terms of a predetermined set of constituents.

In conventional macrorealistic models , the only known tractable way to fulfill the constraints is to elevate the single-particle Lindblad operators into a mass-weighted sum of single-particle operators. Condition (ii) is then automatically fulfilled when the individual operators are chosen as self-adjoint. Condition (iii) follows if each single-particle summand is weighted in proportion to its mass. The growing sum of single-particle operators will additionally ensure that (iv) the total collapse rate amplifies with the total mass.

Ref.~\cite{DCSL} proposes to follow the same ansatz for the dCSL model and subject an $N$-particle state $\rho$ to a dissipator generated by the sum of the single-particle Lindblad operators \eqref{eq:dCSL_K},
\begin{equation} \label{eq:dCSL_ME_Npart_ansatz}
    \frac{\diff}{\diff t} \rho = -\frac{i}{\hbar} \left[ \oH,\rho \right] + \int\diff^3 \vy \, \cD \left[ \sum_{n=1}^N \sqrt{\Gamma_{n}}\oK_{n}(\vy) \right] \rho.
\end{equation}
We allow the measurement rates and the other free parameters in the single-particle operators acting on each individual constituent to depend on its mass $m_n$,
\begin{equation}\label{eq:dCSL_Kn}
    \oK_n (\vy) = \left( \frac{\alpha_n}{2\pi\hbar} \right)^{3/2} \int\diff^3 \vq \, \e^{i\alpha_n \vq \cdot (\vy-\ovx_n)/\hbar} \sqrt{\mu_n (\ovp_n - \vq)} .
\end{equation}
Identical quantum particles of equal mass share the same $\Gamma_n$, $\alpha_n$, and $\mu_n$, which renders the ansatz \eqref{eq:dCSL_ME_Npart_ansatz} exchange symmetry-preserving. Note that we do not yet need to restrict to the Gaussian measurement distribution in \eqref{eq:dCSL_mu_alpha}. 

The major shortcoming of this many-body dCSL ansatz follows from the feedback effect, which renders the single-particle operators \eqref{eq:dCSL_Kn} not self-adjoint: Condition (ii) does not hold, and if we trace over $k$ particles, the dCSL dissipator in \eqref{eq:dCSL_ME_Npart_ansatz} will not retain its form restricted to the $N-k$ remaining ones. Although there is no direct interaction mediated by a potential, the vicinity to the $k$ neglected masses can now influence the state of the others through the dCSL mechanism. The rather technical proof of this observation is given in Appendix \ref{App_innocent_bystander}.

Regarding condition (iii), consider a point-like compound particle in which the $N$ constituents are rigidly bound close to the center of mass such that their relative motion is not resolved by the dCSL position measurement. We may then approximate each constituent position and momentum by the center-of-mass coordinates, $\ovx_n \approx \ovx_{\rm c.m.} = \sum_k \ovx_k/M$ and $\ovp_n \approx m_n \ovp_{\rm c.m.}/M = (m_n/M) \sum_k \ovp_k$. The sum of Lindblad operators in \eqref{eq:dCSL_ME_Npart_ansatz} turns into a single operator acting on the center of mass, 
\begin{align} \label{eq:dCSL_K_cm}
    &\sqrt{\Gamma_{\rm c.m.}} \oK_{\rm c.m.} (\vy) \equiv \sum_n \sqrt{\Gamma_n} \oK_n (\vy) \\
    &\approx \sum_n \sqrt{\Gamma_n} \left( \frac{\alpha_n}{2\pi\hbar}\right)^{3/2} \int\diff^3\vq \, e^{i\alpha_n \vq \cdot (\vy-\ovx_{\rm c.m.})/\hbar} \nonumber \\
    &\qquad \times \sqrt{\mu_n \left( \frac{m_n}{M}\ovp_{\rm c.m.}-\vq \right)} \nonumber \\
    &= \sum_n \sqrt{\Gamma_n} \left( \frac{\alpha_n m_n}{2\pi\hbar M}\right)^{3/2} \int\diff^3\vq_n \, e^{i(\alpha_n m_n/M\hbar)\vq_n \cdot (\vy-\ovx_{\rm c.m.})} \nonumber \\
    &\qquad \times \sqrt{ \left(\frac{m_n}{M}\right)^3 \mu_n \left[ \frac{m_n}{M}(\ovp_{\rm c.m.}-\vq_n) \right]}, \nonumber    
\end{align}
where we substituted $\vq_n = M\vq/m_n$. In order to make this a unique operator acting on a test mass $M$, regardless of the chosen subdivision into constituent masses $m_n$, we demand that each term under the $n$-sum be proportional to $m_n$. This is achieved by scaling the individual  $\Gamma_n$, $\alpha_n$, and $\mu_n$ with respect to fixed reference parameters $\Gamma_0$, $\alpha_0$, and $\mu_0$ at the chosen reference mass $m_0$,
\begin{align}
    \Gamma_n &\equiv \Gamma(m_n) = \frac{m_n^2}{m_0^2}\Gamma_0, \qquad \alpha_n \equiv \alpha (m_n) = \frac{m_0}{m_n}\alpha_0, \nonumber \\
    \mu_n (\vq) &\equiv \mu(\vq;m_n) = \frac{m_0^3}{m_n^3} \mu_0 \left( \frac{m_0}{m_n} \vq \right) .
    \label{eq:dCSL_mass_scaling_law}
\end{align}
Inserting these expressions into \eqref{eq:dCSL_K_cm}, one immediately obtains that dCSL acts on the point-like compound's center of mass consistently like on a single point particle of mass $M$.

An important consequence of the mass scaling is that the $m_0$ can no longer be regarded as merely an arbitrary choice of reference mass units. The friction rate $\alpha(m)$ increases beyond the reference value $\alpha_0$ for smaller masses $m<m_0$ until it may eventually exceed the critical value $\alpha(m) \geq 2$ at which friction turns into heating. The dCSL model should no longer be deemed valid below this critical mass scale. Given that dCSL is a nonrelativistic model and thus only applicable from the atomic scale onward, the conventional value $m_0 = 1\,$u and any $\alpha_0 \ll 2$ would be a safe choice.

Rephrasing the scaling rules \eqref{eq:dCSL_mass_scaling_law} via \eqref{eq:dCSL_mu_alpha} in terms of the dCSL parameters $k$ and $r_{\rm CSL}$ defined in the literature, we arrive at a nonlinear mass scaling with respect to the reference values $k_0$ and $r_0$ at $m_0$ ,
\begin{align}
    k(m) &= \frac{m_0/m}{1 + \left(1- m_0/m\right)k_0} k_0, \nonumber \\
    r_{\rm CSL} (m) &= \left[ 1 + \left(1- \frac{m_0}{m}\right)k_0 \right] r_0.
\end{align}
In Ref.~\cite{DCSL}, the dCSL model was introduced with a universal (mass-independent) length scale parameter $r_{\rm CSL}$ and a friction parameter that is linear in mass, $k\propto m$. This would obey the consistent scaling requirement (iii) only approximately, in the limit $k_0 \ll 1$ and for superatomic masses $m \geq m_0$.

In summary, a straightforward and scale-invariant many-body extension of the dCSL model based on the operator sum ansatz of frictionless models fails to fulfill the usual  consistency requirements. A recent alternative approach to incorporate friction into collapse models could circumvent this problem \cite{dibartolomeo2023linear}.

\section{Conclusions} \label{sec:conclusions}

We presented a versatile and intuitive master equation framework to describe friction forces in the quantum domain. Friction models are based on random repeated momentum measurements of arbitrary resolution, followed by unitary momentum displacements of arbitrary strength as feedback. They make a quantum particle equilibrate to a finite-energy steady state that is, however, generally not thermal. 

The model master equations with their bounded Lindblad operators could be realized, mathematically or in practice, as the ensemble averages of continuous diffusion or piecewise deterministic Poisson processes. Moreover, in the diffusive limit of weak measurements and feedback, they reduce to the familiar Caldeira-Leggett form with unbounded Lindblad operators linear in position and momentum. 

For the specific case of linear friction, we showed an equivalent representation in terms of random position measurements followed by momentum squeezing and linear translations. This uncovered a link to objective collapse theories: the dissipative extension of the continuous spontaneous localization model \cite{DCSL} can be viewed as a representative of our measurement-feedback framework. We found that its many-particle generalization comes with previously unnoticed  technical drawbacks such as the possible buildup of correlations between non-interacting test masses via the feedback force. This calls for a closer inspection of the observable consequences of spontaneous collapse models with friction or with other feedback forces such as gravity \cite{PhysRevD.93.024026, khosla2018classical,carney2023strongly}.

Future work could tap into possible experimental implementations of non-diffusive measurement-feedback friction, compare the dynamics of diffusive and non-diffusive friction models, and study nonlinear feedback forces. 

\acknowledgements

MG acknowledges support from the House of Young Talents of the University of Siegen. We thank Max Weinberg for a valuable mathematical hint.

\onecolumngrid
\appendix


\section{Action of the adjoint channel}
\label{App_adj_channel}

Here we prove the expressions \eqref{eq:Lambda_F} and \eqref{eq:Lambda_G} for the action of the adjoint measurement-feedback channel \eqref{eq:Lambda_adj} on arbitrary operator-valued functions $F(\ovx)$ and $G(\ovp)$. To this end, we simplify the expectation values of these functions with respect to the transformed state $\Lambda (\rho)$. For $F$, we get

\begin{align}
    \left\langle F(\ovp) \right\rangle_{\Lambda [\rho]} &= \tr \left\{ F(\ovp) \ \Lambda \left[ \rho \right]  \right\}  = 
    \int \diff^3 \vq \ \tr \left\{ F(\ovp) \ \oU (\vq) \oM(\vq) \ \rho \  \oM(\vq) \oU ^\dagger(\vq)  \right\} =
    \int \diff^3 \vq \ \tr \left\{ F\left( \ovp -  \vf(\vq)\right) \  \mu\left( \ovp-\vq \right) \rho  \right\} \nonumber \\ 
    &= 
    \int \diff^3 \vp \ \bra{\vp} \rho \ket{\vp}  \int \diff^3 \vq \ F \left( \vp - \vf(\vq) \right) \mu \left( \vp - \vq \right) = 
    \int \diff^3 \vp \ \bra{\vp} \rho \ket{\vp}  \int \diff^3 \vQ \ F\left( \vp - \vf \left( \vp - \vQ \right) \right) \mu \left( \vQ \right) \nonumber \\ 
    &= \int \diff \vp \ \bra{\vp} \rho  \erw_{\mu(\vQ)} \left[ F \left( \ovp -  \vf\left( \ovp - \vQ \right) \right) \right] \ket{\vp} = \left\langle \erw_{\mu(\vQ)} \left[ F \left( \ovp -  \vf\left( \ovp - \vQ \right) \right) \right] \right\rangle_{\rho},
\end{align}
where we used the cyclic property of the trace, applied the momentum kick operators on $F$, expanded the trace in the momentum representation, and substituted $\vQ(\vq) = \vp - \vq$. Since the result must holds for all states $\rho$, we can identify the action of the adjoint channel $\Lambda^\da$ as \eqref{eq:Lambda_F}.

Similarly, we arrive at \eqref{eq:Lambda_G} for $G$ by simplifying

\begin{align}
    \left\langle G(\ovx) \right\rangle_{ \Lambda \left[\rho \right]} &= \tr \left\{ G(\ovx) \Lambda \left[ \rho \right] \right\} = 
    \int \diff^3 \vq \ \tr \left\{ G(\ovx) \oM(\vq) \rho \oM(\vq)  \right\} = 
    \int \diff^3 \vy \  \nu(\vy) \tr \left\{ \e^{-i \frac{\vy \cdot \ovp}{\hbar}} G(\ovx) \e^{+i \frac{\vy \cdot \ovp}{\hbar}} \ \rho  \right\} \nonumber \\ 
    &= 
    \int \diff^3 \vy \ \nu(\vy) \left\langle G(\ovx - \vy) \right\rangle_{\rho} =
    \left\langle \erw_{\nu(\vy)} \left[  G(\ovx - \vy)  \right] \right\rangle_{\rho}.
\end{align}
Here we have made use of the identity \eqref{eq_random_trans}.

\section{Diffusion limit}
\label{App_diff_limit}

Here we calculate the diffusive expansion of the measurement-feedback channel $\Lambda$ in \eqref{eq:Lambda_channel} and the associated master equation \eqref{eq_master_general} up to second order in the position and momentum quadratures. This is valid for states $\rho$ restricted to much smaller momenta than what the measurement distribution $\mu$ can resolve, and for feedback kick strengths much smaller than the characteristic momentum scale occupied by $\rho$.

In order to simplify notation, we set $\vQ = - \vq$, introduce the Hesse matrix $\Hess_{\mu (\vQ)}$ with elements $[ \Hess_{\mu (\vQ)} ]_{jk} = \partial^2 \mu (\vQ)/\partial Q_j \partial Q_k$, and use the '$\approx$' sign whenever we neglect third-order terms in the components of $\ovx$ and $\ovp$. Moreover, we only consider asymptotically well-behaved distributions $\mu\neq 0$ that obey  $\vert \vq \vert^2 \mu(\vq) \to 0 $ , $\vert \vq \vert^2 \vert \nabla \mu(\vq) \vert  \to 0 $, and $\vert \vq \vert^2 \vert  \vf(\vq) \nabla \mu(\vq) \vert  \to 0$ for $\vert \vq \vert \to \infty$. We first expand the term
\begin{equation}
    \sqrt{\mu\left( \ovp - \vq \right) } = \sqrt{\mu\left( \vQ + \ovp \right)} \approx \sqrt{\mu\left( \vQ \right)} 
    + \frac{\ovp \cdot \nabla \mu(\vQ) }{2 \sqrt{\mu(\vQ)}} + \frac{1}{4} \left[ \frac{\ovp^\trans \Hess_{\mu ( \vQ )} \ovp}{\sqrt{\mu(\vQ)}} - \frac{\left( \ovp \cdot \nabla \mu(\vQ) \right)^2}{2 \mu^{3/2}(\vQ)} \right]. 
\end{equation}
This yields the unnormalized post-measurement state 
\begin{align}
     \sigma_{\vQ} &\equiv \sqrt{\mu\left( \vQ + \ovp \right)} \ \rho \  \sqrt{\mu\left( \vQ + \ovp \right)}
     \begin{aligned}[t] 
     \ \approx \ & 
     \mu\left( \vQ \right) \rho + \frac{1}{2} \left\{ \ovp \cdot \nabla \mu \left( \vQ \right), \rho \right\} 
     +
     \frac{1}{4} \left\{ \ovp^\trans \ \Hess_{\mu ( \vQ )} \ovp , \rho \right\} \\ 
     & + 
     \frac{\left( \ovp \cdot \nabla \mu(\vQ) \right) \rho \left( \ovp \cdot \nabla \mu(\vQ) \right)}{4 \mu(\vQ)} - \frac{1}{8} \left\{ \frac{(\ovp \cdot \nabla \mu(\vQ))}{ \mu(\vQ)}, \rho  \right\} 
     \end{aligned}
     \nonumber \\ 
     &= \mu\left( \vQ \right) \rho + \frac{1}{2} \left\{ \ovp \cdot \nabla \mu \left( \vQ \right), \rho \right\} 
     + \frac{1}{4} \left\{ \ovp^\trans \ \Hess_{\mu ( \vQ )} \ovp , \rho \right\} - \frac{1}{8 \mu(\vQ)} \left[ \ovp \cdot \nabla \mu(\vQ) ,  \left[ \ovp \cdot \nabla \mu(\vQ), \rho  \right] \right]. 
\end{align}
For feedback kicks of the form $\vf(\vq) = \bar{f}(q) \vq/q = -\vf (\vQ) $, we expand the feedback unitary applied to the post-measurement state as
\begin{align}
    \e^{i \vf(\vQ) \cdot \ovx /\hbar} \sigma_{\vQ} \e^{- i \vf(\vQ) \cdot \ovx /\hbar} &\approx \left( \mathds{1} + \frac{i}{\hbar} \left( \vf(\vQ) \cdot \ovx \right) - \frac{1}{2 \hbar^2} \left( \vf(\vQ) \cdot \ovx \right)^2 \right) \sigma_{\vQ} \left( \mathds{1} - \frac{i}{\hbar} \left( \vf(\vQ) \cdot \ovx \right) - \frac{1}{2 \hbar^2} \left( \vf(\vQ) \cdot \ovx \right)^2 \right) \nonumber \\ 
    &\approx \sigma_{\vQ} + \frac{i}{\hbar} \left[ \vf(\vQ) \cdot \ovx , \sigma_{\vQ} \right]
    +
    \frac{1}{\hbar^2} (\vf(\vQ) \cdot \ovx) \ \sigma_{\vQ} \ (\vf(\vQ) \cdot \ovx) - \frac{1}{2\hbar^2} \left\{ (\vf(\vQ) \cdot \ovx)^2, \sigma_{\vQ} \right\} \nonumber \\ 
    &= \sigma_{\vQ} + \frac{i}{\hbar} \left[ \vf(\vQ) \cdot \ovx , \sigma_{\vQ} \right] 
    - \frac{1}{2 \hbar^2} \left[ \vf(\vQ) \cdot \ovx  , \left[ \vf(\vQ) \cdot \ovx , \sigma_{\vQ} \right]\right]. 
\end{align} 
With the above expression $\sigma_{\vQ}$ for the post-measurement state, we obtain to second order,
\begin{align}
    &\e^{i \vf(\vQ) \cdot \ovx /\hbar} \sqrt{\mu\left( \vQ + \ovp \right)} \ \rho \  \sqrt{\mu\left( \vQ + \ovp \right)} \ \e^{- i \vf(\vQ) \cdot \ovx /\hbar} \nonumber  \\ 
    &\approx \mu\left( \vQ \right) \rho + \frac{1}{2} \left\{ \ovp \cdot \nabla \mu \left( \vQ \right), \rho \right\} 
     + \frac{1}{4} \left\{ \ovp^\trans \ \Hess_{\mu ( \vQ )} \ovp , \rho \right\}  
      - \frac{1}{8 \mu(\vQ)} \left[ \ovp \cdot \nabla \mu(\vQ) ,  \left[ \ovp \cdot \nabla \mu(\vQ), \rho  \right] \right] \nonumber \\ 
      &\quad + \frac{i}{\hbar} \mu(\vQ) \left[ \vf(\vQ) \cdot \ovx, \rho  \right] 
       +  \frac{i}{2\hbar} \left[ \vf(\vQ) \cdot \ovx, \left\{ \ovp \cdot \nabla \mu(\vQ), \rho \right\}  \right] 
       - \frac{1}{2\hbar^2} \mu(\vQ) \left[ \vf(\vQ) \cdot \ovx  , \left[ \vf(\vQ) \cdot \ovx , \rho \right]\right]. 
\end{align}
The transformed state $\Lambda[\rho]$ follows after integration over $\vQ$, which can be done on each term individually. 
The second and third terms vanish by virtue of Gauss' theorem. For example, making use of the fact that $\vert \vQ \vert^2 \mu(\vQ) \to 0$, we can deduce

\begin{equation}
    \label{eq_surface_trick}
    \left\vert \int_{|\vQ| \leq R} \diff^3 \vQ \  \nabla \mu(\vQ) \right\vert = \left\vert \int_{|\vQ|=R} \diff^2 \vQ \ \mu(\vQ) \right\vert \leq 4 \pi R^2 \sup_{\vert \vQ \vert = R} \mu(\vQ) \xrightarrow[R \to \infty]{} 0, 
\end{equation}
so that $\int\diff^3 \vQ \, \nabla \mu (\vQ) = 0$. Similarly, since $[\Hess_{\mu(\vQ)}]_{jk} = [\nabla \partial_j\mu]_k$, we also find that $\int\diff^3 \vQ \, \Hess_{\mu (\vQ)} = 0$. 

Hence the transformed state reads as
\begin{align}
    \Lambda[\rho] &= \int_{\reel^3} \diff^3 \vQ \ \e^{i \vf(\vQ) \cdot \ovx /\hbar} \sqrt{\mu\left( \vQ + \ovp \right)} \ \rho \  \sqrt{\mu\left( \vQ + \ovp \right)} \ \e^{- i \vf(\vQ) \cdot \ovx /\hbar} \nonumber \\ 
    &= \rho + \frac{i}{\hbar} \left[ \erw_{\mu(\vQ)} \left[ \vf(\vQ) \right] \cdot \ovx, \rho \right] 
    - \frac{1}{8} \sum_{j,k} \underbrace{ \int \diff^3 \vQ \frac{1}{\mu(\vQ)} \frac{\partial \mu(\vQ)}{\partial Q_j} \frac{\partial \mu(\vQ)}{\partial Q_k}}_{\equiv \tilde{A}_{jk}} \left[ \op_j, \left[ \op_k, \rho \right] \right]
    \nonumber \\ 
    &\quad + \frac{i}{2 \hbar} \sum_{j,k} \underbrace{ \int \diff^3 \vQ \ f_j (\vQ) \frac{\partial \mu(\vQ)}{\partial Q_k} }_{\equiv \tilde{B}_{jk}} \left[ \ox_j, \left\{ \op_k, \rho \right\} \right] - \frac{1}{2 \hbar^2} \sum_{j,k} \underbrace{ \int \diff^3 \vQ \ f_j (\vQ) f_k (\vQ) \mu(\vQ)}_{\equiv \tilde{C}_{jk}} \left[ \ox_j, \left[ \ox_k, \rho \right] \right] . \label{eq:Lambda_diffExpansion}
\end{align}
The auxiliary matrices defined here can be rewritten in a compact manner as the expectation values
\begin{equation}
     \tilde{A}_{jk} = - \erw_{\mu} \left[ \partial_k \frac{\partial_j \mu}{\mu} \right] = - \erw_{\mu} \left[ \partial_k \partial_j \ln(\mu) \right],\qquad \tilde{B}_{jk} = - \erw_{\mu} \left[ \partial_k f_j  \right], \qquad \tilde{C}_{jk} = \erw_{\mu} \left[ f_j f_k \right].
\end{equation}
For the first two matrices, we integrated by parts assuming once again vanishing boundary terms. 

The matrix $\tilde{A}$ can also be expressed in terms of moments of the conjugate distribution $\nu(\vy)$. To this end, we first notice 
\begin{equation}
 \frac{\partial_j \mu(\vQ)}{2\sqrt{\mu(\vQ)}} = \partial_j \sqrt{\mu(\vQ)} = -\frac{i}{\hbar} \int  \frac{\diff^3 \vy}{\sqrt{2 \pi \hbar}^3} \ \e^{-i \vQ \cdot \vy/\hbar} y_j \sqrt{\nu(\vy)}.
\end{equation}
Consequently, we can rewrite
\begin{align}
    \tilde{A}_{jk} &= 4 \int \diff^3 \vQ \frac{\partial_j \mu(\vQ)}{2\sqrt{\mu(\vQ)}} \frac{\partial_k \mu(\vQ)}{2\sqrt{\mu(\vQ)}} =  
    -\frac{4}{\hbar^2}\int \frac{\diff^3 \vQ}{(2 \pi \hbar)^3}   \int \diff^3 \vy \int \diff^3 \vz \ \e^{-i \vQ \cdot (\vy + \vz)/\hbar} \  y_j z_k \sqrt{\nu(\vy)} \sqrt{\nu(\vz)} \nonumber \\  
    &= - \frac{4}{\hbar^2} \int \diff^3 \vy \  y_j (-y_k) \nu(\vy) 
    = \frac{4}{\hbar^2} \erw_{\nu(\vy)} \left[y_j y_k \right].
\end{align}
The diffusion matrices \eqref{eq:diffABC} in the main text are simply given by $A = \tilde{A}/8$, $B = \tilde{B}/2\hbar$, and $C = \tilde{C}/2\hbar^2$, so that the second order expansion of the channel assumes the compact form

\begin{align}
    \Lambda[\rho] = \rho + \frac{i}{\hbar} \left[ \erw_\mu \left[ \vf \right] \cdot \ovx , \rho \right] 
    - \sum_{j,k} \left(
    A_{jk} \left[ \op_j \left[ \op_k, \rho \right] \right]
    + 
    iB_{jk} \left[ \ox_j, \left\{ \op_k,  \rho \right\} \right] 
    + 
    C_{jk} \left[ \ox_j \left[ \ox_k, \rho \right] \right] \right).
\end{align}


\section{Transformation from momentum to position measurement under linear force}
\label{App_mom_to_pos}

Here we show the equivalence of the two representations \eqref{eq:Lambda_L_K} of the linear friction measurement-feedback channel, by deriving the explicit form of the alternative Kraus operators \eqref{eq:Ky} from the original Kraus operators, 

\begin{equation}
\oL\left( \vq \right) =  \e^{-i \alpha\vq \cdot \ovx/\hbar} \sqrt{\mu \left( \ovp - \vq \right)} = \sqrt{\mu \left( \ovp - \left( 1 - \alpha \right)\vq \right)} \e^{-i \alpha \vq \cdot \ovx / \hbar} \ .
\end{equation}
The alternative operators \eqref{eq:KfromL} are defined as the Fourier transform $\oK  ( \vy ) = \sqrt{\alpha/2 \pi \hbar}^3 \int \diff^3 \vq \ \e^{i \alpha \vq \cdot \vy/\hbar}  \oL ( \vq ) $ for $\alpha > 0$, which we can simplify by expanding in momentum and position eigenstates to the left and to the right of the operator, respectively,
\begin{align}
    \oK  \left( \vy \right) &= \int\diff^3\vp \int\diff^3\vx \, |\vp\ra\la\vp|\oK(\vy)|\vx\ra\la \vx| = \frac{\alpha^{3/2}}{(2 \pi \hbar)^3} \int \diff^3 \vp \int \diff^3 \vx \, |\vp\ra \la \vx| \e^{-i\vp\cdot\vx/\hbar} \int\diff^3\vq \,
  \e^{i \alpha \vq \cdot (\vy - \vx)/\hbar} \sqrt{\mu \left( \vp - \left( 1 - \alpha \right)\vq \right)} \nonumber \\
  &= \frac{\alpha^{3/2}}{(2 \pi \hbar)^3} \int \diff^3 \vp \int \diff^3 \vx \, |\vp\ra \la \vx| \e^{-i\vp\cdot\vx/\hbar} \int \frac{\diff^3\vQ}{|1-\alpha|^3} \sqrt{\mu(\vQ)} \e^{-i \alpha (\vp - \vQ) \cdot (\vx - \vy)/(1-\alpha)\hbar }  \nonumber \\
    &= \frac{\alpha^{3/2}}{(2 \pi \hbar)^{3/2}} 
    \int \diff^3 \vp \, \e^{i \alpha \vp  \cdot \vy/(1-\alpha)\hbar } \int \diff^3 \vx \ \ket{\vp} \braket{\vp \big\vert \frac{\vx}{1- \alpha}} \bra{\vx}
    \frac{(2\pi\hbar)^{3/2}}{\vert 1- \alpha \vert^3} \sqrt{ \nu \left[ \frac{\alpha}{1-\alpha}(\vx-\vy)\right]} \nonumber \\ 
    &= \left( \frac{\alpha}{ \vert 1-\alpha \vert}\right)^{3/2} \e^{i \alpha \ovp  \cdot \vy/(1-\alpha)\hbar} \underbrace{
    \int \frac{\diff^3 \vx}{ \vert 1-\alpha \vert^{3/2}} \ket{\frac{\vx}{1- \alpha}} \bra{\vx}}_{ \equiv \oW(\alpha)}
    \sqrt{ \nu \left[ \frac{\alpha}{1-\alpha}(\vx-\vy)\right]}. \label{eq_app_pos_channel} 
\end{align}
In the second line, we substituted $\vQ(\vq) = \vp - (1-\alpha) \vq$, assuming $\alpha \neq 1$. The third line uses the Fourier representation \eqref{eq:sqrt_nu} of the conjugate function $\sqrt{\nu}$.
We distinguish the two cases $0<\alpha<1$ and $\alpha >1$. In the former case, which is also given in \eqref{eq:Ky} in the main text, the auxiliary operator $\oW(\alpha)$ is a scaling transformation, 
\begin{equation}
    \oW(\alpha) = \oS(\alpha) = \exp \left[ \frac{i \ln  \vert 1-\alpha \vert }{2 \hbar} \left\{ \ovx, \ovp \right\} \right] = \bigotimes_{j=1}^3 \exp \left[ \frac{i \ln  \vert 1-\alpha \vert }{2 \hbar} \left\{ \ox_j, \op_j \right\} \right]\ . \label{eq_App_Scaling_def}
\end{equation}
It describes (anti-)squeezing of the position (momentum) quadratures, since
\begin{equation}
    \oS^\dagger( \alpha) \ovx \oS(\alpha) = \frac{1}{\vert 1 - \alpha \vert}  \ovx, \qquad \oS^\dagger(\alpha)\ovp \oS(\alpha) = \vert 1 - \alpha \vert \ovp, \qquad 
    \oS \left( \alpha \right) \ket{\vx} = \frac{1}{\vert 1 - \alpha \vert^{3/2} } \ket{ \frac{\vx}{\vert 1 - \alpha \vert}} \ . 
\end{equation}
Introducing bosonic ladder operators $\oa_j$ with respect to an arbitrary harmonic frequency $\omega$, as done in the main text, we can rewrite $\{ \ox_j,\op_j \} = -i\hbar (\oa_j^2 - \oa_j^{\da 2})$, which yields the textbook definition of a squeezing operator \eqref{eq:S_squeezingForm} with squeezing parameter $\xi_j = \ln |1-\alpha|$ \cite{gerry_knight}.

For $\alpha>1$, an additional parity operation must be performed, $\hat{\mathcal{P}} \ket{\vx} = \ket{- \vx}$, 
\begin{align}
    \oW(\alpha) = \int \frac{\diff^3 \vx}{ \vert 1-\alpha \vert^{3/2}} \ket{\frac{\vx}{1- \alpha}} \bra{\vx} = \int \frac{\diff^3 \vx}{ \vert 1-\alpha \vert^{3/2}} \ket{ \frac{ - \vx}{ \vert 1- \alpha \vert}} \bra{\vx} = \hat{\mathcal{P}} \ \oS(\alpha).
\end{align}
What remains is the special case $\alpha=1$ for which the squeezing transformation and the feedback translation become singular. From the first line in \eqref{eq_app_pos_channel}, we can directly deduce a representation of the channel in terms of singular sharp position measurements, $\oK(\vy) = (2 \pi \hbar)^{3/2} \sqrt{\mu(\ovp)} \delta ( \ovx - \vy ) = (2 \pi \hbar)^{3/2} \sqrt{\mu(\ovp)} |\vy \ra\la\vy|$.

Here, the position measurement is not followed by outcome-dependent feedback operation, but instead by the unsharp momentum measurement $\sqrt{\mu(\ovp)}$ that regularizes the post-measurement state.

\section{Transformation of the characteristic function under linear friction}
\label{App_trafo_char}

In order to study the fix point of the linear-friction channel, we first calculate the transformation \eqref{eq:charFunc_trafo} of the characteristic function corresponding to the transformation $\rho \to \Lambda[\rho]$,
\begin{align}
    &\chi_{\Lambda [\rho]} \left( \vP, \vX \right) 
    = \tr \left\{ \e^{i \left( \vP \cdot \ovx + \vX \cdot \ovp \right)/\hbar} \Lambda \left[ \rho \right] \right\}
    = \e^{-i  \vP \cdot \vX/2 \hbar} \int \diff^3 \vp \ \e^{i \vX \cdot \vp/ \hbar} \bra{\vp - \vP} \Lambda \left[ \rho \right] \ket{\vp} \nonumber \\ 
    &= \e^{-i \vP \cdot \vX/2 \hbar} \int \diff^3 \vp \ \int \diff^3 \vq \  \e^{i \vX \cdot \vp /\hbar} \bra{\vp + \alpha \vq -  \vP} \sqrt{\mu\left( \ovp - \vq \right)} \rho \sqrt{\mu\left( \ovp - \vq \right)} \ket{\vp + \alpha \vq} \nonumber \\ 
    &= \e^{-i \vP \cdot \vX/2 \hbar} \int \diff^3 \vR \int \diff^3 \vQ \ \e^{i \vX \cdot \left[ (1-\alpha) \vR + \alpha \vQ \right] / \hbar} \sqrt{\mu \left(\vQ - \vP \right) \mu \left( \vQ \right) } \bra{\vR - \vP} \rho \ket{\vR} \nonumber \\ 
    &= \underbrace{\left( \int \diff^3 \vQ \ \e^{i\alpha \vX \cdot \vQ/ \hbar} \sqrt{\mu \left(\vQ - \vP\right) \mu \left( \vQ \right) } \right)}_{=A(\vP, \vX)} \e^{-i  \alpha \vP \cdot \vX/2 \hbar} \underbrace{ \e^{-i (1-\alpha) \vP \cdot \vX /2 \hbar}
    \int \diff^3 \vR \ \e^{i (1-\alpha) \vX \cdot \vR /\hbar} \bra{\vR - \vP} \rho \ket{\vR}}_{=\chi_{\rho}(\vP, (1-\alpha)\vX)} \nonumber \\ 
    &=A(\vP, \vX) \e^{-i \alpha \vP \cdot \vX/2 \hbar} \chi_{\rho}(\vP, (1-\alpha) \vX), \label{eq_App_char_trafo}
\end{align}
In the third line, we substituted 
\begin{equation}
   \left( \begin{array}{cc}
         \vR  \\
         \vQ
    \end{array} \right) 
    = 
   \left(  \begin{array}{cc}
        1 & \alpha  \\
        1 & \alpha - 1  
    \end{array} \right) 
    \left(  \begin{array}{cc}
         \vp  \\
          \vq
    \end{array}
    \right).
\end{equation}
From the above result \eqref{eq_App_char_trafo} follows the condition \eqref{eq:fixPointCond} for states to be a fix point of $\Lambda$, $\chi_{\Lambda[\rho]} = \chi_{\rho}$. Taking the absolute value of that condition, $|\chi_{\rho } ( \vP, ( 1 - \alpha  ) \vX )| =|\chi_{\rho } ( \vP, \vX )| / |A ( \vP, \vX )|$, and iterating this relation $n$ times, we obtain
\begin{equation}
     \left\vert \chi_{\rho } \left( \vP, \left( 1 - \alpha  \right)^n \vX \right) \right\vert  = \frac{\left\vert \chi_{\rho } \left( \vP, \vX \right) \right\vert}{\prod_{j=0}^{n-1} \left\vert A \left( \vP, (1-\alpha)^j \vX \right) \right\vert } \label{eq_iterated_char} .
\end{equation}
In the denominator, we can upper-bound each term by unity with help of the triangle and the Cauchy-Schwarz inequality,
\begin{equation}
    \left\vert A \left( \vP, \vX \right) \right\vert \leq A \left( \vP, 0 \right) = \int \diff^3 \vq \ \sqrt{ \mu \left( \vq - \vP \right) \mu \left( \vq \right)} \leq 1 \ , 
\end{equation}
Note that the bound is saturated if and only if  $\sqrt{\mu \left(\vq \right)}$ and $\sqrt{\mu \left(\vq - \vP \right)}$ are linearly dependent functions of $\vq$. Due to the normalization, the only possibility is $\mu \left( \vq \right) = \mu \left( \vq - \vP \right) $ for all $\vq \in \reel^3$, which implies $\vP = 0$. Hence, $ | A ( \vP, \vX ) | \leq |A(\vP,0)| < 1 $ for any $\vP \neq 0$, so that \eqref{eq_iterated_char} results in 
\begin{equation}
    \left\vert \chi_{\rho } \left( \vP, \left( 1 - \alpha  \right)^n \vX \right) \right\vert \geq \underbrace{ \left\vert A \left( \vP, 0 \right) \right\vert^{-n}}_{\xrightarrow[n \to \infty]{} \infty} \left\vert \chi_{\rho } \left( \vP, \vX \right) \right\vert \ . \label{eq:chi_singularity_statement}
\end{equation}
Since $(1-\alpha)^n \vX \xrightarrow[n \to \infty]{} 0$ for any $\alpha \in (0,2)$ that leads to friction, we conclude that the fix point must correspond to a singular characteristic function: If we had a proper state with a finite, normalized Wigner function, then its Fourier transform $\chi_\rho$ would be a finite function with $\chi_\rho (0,0)=1$ and a finite support. However, for any pair $\vP \neq 0$, $\vX \neq 0$ with $\chi_{\rho } \left( \vP, \vX \right) \neq 0$, \eqref{eq:chi_singularity_statement} implies that $|\chi_{\rho } \left( \vP, 0 \right)| \to \infty$, i.e., a singular point at $\left( \vP, 0 \right)$. The corresponding Wigner function would no longer be normalizable.

\section{Rotational covariance of the measurement feedback channel}
\label{App_Rot_cov}

Here we prove the equivalence between the rotation covariance of the measurement-feedback channel $\Lambda$ and the isotropy of the measurement distribution $\mu$, under the assumption of a central feedback force, $\vf(\vq) = \bar{f}(q)\vq/q$.
Consider the unitary representation of the rotation group, $\oU_R$ for $R \in \mathrm{SO(3)}$ with $\oU_R \ket{\vx} = \ket{R \vx}$, $\oU_R \ovx \oU_R^\dagger = R^{\mathrm{T}} \ovx $, and $\oU_R \ovp \oU_R^\dagger = R^{\mathrm{T}} \ovp$. 
Applying such a unitary to the transformed state $\Lambda[\rho]$ yields
\begin{align}
    \oU_R \Lambda \left[ \rho \right] \oU_R^\dagger &= \int \diff^3 \vq  \left[ \oU_R  \e^{-i \bar{f}(q) \vq\cdot\ovx/\hbar q} \oU_R^\dagger \right]
     \left[ \oU_R \sqrt{\mu \left( \ovp - \vq \right)} \oU_R^\dagger \right] \left[ \oU_R \rho \oU_R^\dagger \right] \left[ \oU_R \sqrt{\mu \left( \ovp - \vq \right)} \oU_R^\dagger \right] \left[\oU_R  \e^{i \bar{f}(q) \vq\cdot\ovx/\hbar q} \oU_R^\dagger \right] \nonumber \\
     &= \int \diff^3 \vq \,  \e^{-i \bar{f}(q) \vq\cdot R^\trans \ovx/\hbar q} \,
     \sqrt{\mu \left( R^\trans\ovp - \vq \right)} \, \oU_R \rho \oU_R^\dagger \, \sqrt{\mu \left( R^\trans \ovp - \vq \right)} \, \e^{i \bar{f}(q) \vq\cdot R^\trans \ovx/\hbar q} \nonumber \\
     &= \int \diff^3 \vQ \,  \e^{-i \bar{f}(Q) \vQ\cdot \ovx/\hbar Q} \,
     \sqrt{\mu \left[ R^\trans (\ovp - \vQ) \right]} \, \oU_R \rho \oU_R^\dagger \, \sqrt{\mu \left[ R^\trans (\ovp - \vQ) \right]} \, \e^{i \bar{f}(Q) \vQ\cdot \ovx/\hbar Q},
\end{align}
where we have substituted $\vQ = R\vq$. Rotational covariance of $\Lambda$ means that $\oU_R\Lambda[\rho]\oU_R^\da = \Lambda[\oU_R\rho\oU_R^\da]$ for any $R,\rho$. We see that this is the case if and only if the measurement distribution is isotropic, $\mu(R\vq) = \mu(\vq) \equiv \bar{\mu} (q)$.

\section{dCSL-mediated influence between particles}
\label{App_innocent_bystander}

Many-particle versions of spontaneous collapse models adhere to the principle that the collapse mechanism acts separately on independent, non-interacting particles; their reduced states do not influence each other. This is no longer the case for collapse models with feedback, such as the dCSL model.

To show this, it is sufficient to consider the simplest case of two particles. Let $\Gamma_n$ and $\oK_n (\vy)$ be, respectively, the individual collapse rate and the Lindblad operators associated to the $n$-th particle. Following the sum ansatz \eqref{eq:dCSL_ME_Npart_ansatz} for the two-particle model, the reduced states of the two particles remain independent if
\begin{equation}\label{eq:innocentBystander_condition}
    \tr_2 \left\{ \int \diff^3 \vy \ \mathcal{D}\left[ \sqrt{\Gamma_1} \oK_1(\vy) + \sqrt{\Gamma_2} \oK_2(\vy) \right] \rho \right\} \stackrel{!}{=} \Gamma_1 \int \diff^3 \vy \ \mathcal{D}\left[ \oK_1(\vy) \right] \tr_2 \left\{ \rho \right\}. 
\end{equation}
Under this condition, we retain the single-particle dissipator for the reduced state of, say, the first particle when tracing over the second one.

Let us now insert an arbitrary product state $\rho = \rho_1 \otimes \rho_2$ into the left hand side of \eqref{eq:innocentBystander_condition}, which yields
\begin{align}
    \tr_2 \left\{ \int \diff^3 \vy \ \mathcal{D}\left[ \sqrt{\Gamma_1} \oK_1(\vy) + \sqrt{\Gamma_2} \oK_2(\vy) \right] \rho_1 \otimes \rho_2 \right\} =& \Gamma_1 \int \diff^3 \vy \ \mathcal{D}\left[ \oK_1(\vy) \right]\rho_1 \\ \nonumber
    &+ \frac{1}{2} \sqrt{\Gamma_1 \Gamma_2} \left( \int \diff^3 \vy \tr \left\{ \oK_2^\dagger(\vy) \rho_2 \right \} \left[ \oK_1(\vy), \rho_1 \right] + h.c. \right). \nonumber
\end{align}
Hence, the condition \eqref{eq:innocentBystander_condition} implies that
\begin{equation}
\label{eq_bystander_condition}
 \int \diff^3 \vy \tr \left\{ \oK_2^\dagger(\vy) \rho_2 \right\} \left[ \oK_1(\vy), \rho_1 \right] \stackrel{!}{=}  \int \diff^3 \vy \tr \left\{ \oK_2(\vy) \rho_2 \right\} \left[ \oK_1^\dagger(\vy), \rho_1 \right]
\end{equation}
for any $\rho_1, \rho_2$. This is automatically true for self-adjoint Lindblad operators or operators that obey $\oK_n^\dagger(\vy) = \oK_n^\dagger( - \vy)$, as is the case for collapse models without feedback. 

For the present case with Lindblad operators \eqref{eq:Ky}, consider the improper product state $\rho = \rho_1 \otimes |\vx_2 \ra \la \vx_2|$, which can be arbitrarily well approximated by preparing the second particle in an increasingly sharp wave packet localized at $\vx_2$. The relevant trace expression in \eqref{eq_bystander_condition} then reduces to the real-valued 
\begin{align}
    \tr \left\{ \oK_2(\vy) \rho_2 \right\} &= \bra{\vx_2} \oK_2(\vy) \ket{\vx_2} = \bra{(1-\alpha_2) \vx_2 + \alpha_2 \vy } \sqrt{(1-\alpha_2)^3} 
    \sqrt{\left(\frac{\alpha}{1-\alpha_2} \right)^3 \nu \left[ \frac{\alpha_2}{1-\alpha_2} (\ovx - \vy) \right]} \ket{\vx_2} \nonumber\\ 
    &= \sqrt{\nu \left[ \alpha_2 (\vx_2 - \vy) \right]} \frac{1}{\alpha_2^{3/2}} \delta^{(3)}\left( \vx_2 - \vy \right) 
    = \frac{\sqrt{\nu(0)}}{\alpha_2^{3/2}} \ \delta^{(3)}\left( \vx_2 - \vy \right) = \tr \left\{ \oK_2^\da (\vy) \rho_2 \right\}. \label{eq_bystander_tr}
\end{align}
Therefore, condition \eqref{eq_bystander_condition} requires that  $[\oK_1(\vx_2)-\oK_1^\da (\vx_2),\rho_1] \stackrel{!}{=} 0$ for any chosen $\rho_1$ and $\vx_2$. This is clearly not possible here since the Lindblad operators and their adjoints describe different transformations for $\alpha \neq 0$,
\begin{align}
    \oK_1(\vy) &=  \exp \left( \frac{i\ovp}{\hbar} \cdot \frac{\alpha \vy}{1-\alpha}\right) \oS(\alpha) \nonumber 
    \sqrt{\left(\frac{\alpha}{1-\alpha}\right)^3 \nu \left[ \frac{\alpha}{1-\alpha} (\ovx-\vy)  \right]}, \\ 
    \oK^\dagger_1(\vy) &= \exp \left( -i \alpha  \frac{\vy \cdot \ovp}{\hbar}\right) \oS^\dagger(\alpha) \sqrt{\left(\frac{\alpha}{1-\alpha}\right)^3 \nu \left[ \alpha (\ovx-\vy)  \right]}.
\end{align} 

\end{document}